\documentclass[12pt]{article}

\usepackage[T1]{fontenc}
\usepackage{float}
\usepackage{amsmath}
\usepackage{graphicx}
\usepackage{setspace}
\usepackage[authoryear]{natbib}
\usepackage[latin9]{inputenc}
\usepackage{lmodern}

\usepackage{algpseudocode}
\usepackage{algorithm}

\usepackage{subfig}
\usepackage{amsmath,amssymb,amsthm,setspace,paralist}
\usepackage[colorlinks,citecolor=blue,urlcolor=blue]{hyperref}
\usepackage{epstopdf}
\usepackage{authblk}
\usepackage[normalem]{ulem}

\newcommand{\bsb}{\boldsymbol}
\newcommand{\bm}{\boldsymbol}
\newcommand{\bsbTh}{{\boldsymbol{\Theta}}}
\newcommand{\bsby}{\boldsymbol{y}}
\newcommand{\bsbb}{\boldsymbol{\beta}}
\newcommand{\bsba}{\boldsymbol{a}}
\newcommand{\bsbeps}{\boldsymbol{\epsilon}}
\newcommand{\bsbY}{\boldsymbol{Y}}
\newcommand{\bsbB}{\boldsymbol{B}}
\newcommand{\bsbA}{\boldsymbol{A}}
\newcommand{\bsbC}{\boldsymbol{C}}
\newcommand{\bsbV}{\boldsymbol{V}}
\newcommand{\bsbU}{\boldsymbol{U}}
\newcommand{\bsbD}{\boldsymbol{D}}

\newcommand{\bsbX}{\boldsymbol{X}}
\newcommand{\bsbZ}{\boldsymbol{Z}}

\newcommand{\bsbI}{\boldsymbol{I}}

\newcommand{\bsbmu}{{\boldsymbol{\mu}}}
\newcommand{\bsbx}{\boldsymbol{x}}
\newcommand{\bsbT}{\boldsymbol{T}}
\newcommand{\bsbW}{\boldsymbol{W}}
\newcommand{\bsbSig}{\boldsymbol{\Sigma}}
\newcommand{\bsbv}{\boldsymbol{v}}
\newcommand{\bsbz}{\boldsymbol{z}}
\newcommand{\bsbr}{\boldsymbol{r}}
\newcommand{\bsbt}{\boldsymbol{t}}
\newcommand{\bsbR}{\boldsymbol{R}}

\newcommand{\bsbxi}{\boldsymbol{\xi}}
\newcommand{\bsbXi}{\boldsymbol{\Xi}}
\newcommand{\bsbg}{\boldsymbol{\gamma}}
\newcommand{\bsbSigma}{\boldsymbol{\Sigma}}

\newcommand{\Breg}{{\mathbf{D}}}

\DeclareMathOperator*{\argmin}{argmin}

\DeclareMathOperator{\sgn}{sgn}
\DeclareMathOperator{\vect}{\mbox{vec}\,}
\newcommand{\Proj}{{\mathbf P}}
\newcommand{\avg}{\mathrm{avg}}
\newcommand{\rd}{\,\mathrm{d}}

\theoremstyle{definition}

\theoremstyle{plain}
\providecommand{\examplename}{Example}

\newtheorem{thm}{Theorem}[section]

\newtheorem{lemma}[thm]{Lemma}
\newtheorem{remark}{Remark}

\usepackage{xr}
\begin{document}

\title{On Generalization and Computation of Tukey's Depth: Part I}
\author{Yiyuan She, Shao Tang, and Jingze Liu\\Department of Statistics, Florida State University}%
\date{}
\maketitle

\begin{abstract}
Tukey's depth offers a powerful tool   for nonparametric   inference and estimation,  but       also encounters serious  computational and methodological difficulties   in modern statistical data analysis. This paper studies how to generalize and compute Tukey-type depths    in multi-dimensions. A general framework of   influence-driven    polished subspace depth,   which emphasizes the importance of the underlying influence space and   discrepancy  measure, is introduced.      The  new matrix formulation  enables us to utilize state-of-the-art optimization techniques to develop scalable   algorithms with implementation ease and guaranteed fast convergence.  In particular, half-space depth as well as regression depth can now be computed much faster than previously possible, with the support from  extensive experiments. A companion paper is also offered to the reader in the same issue of this journal. 

\end{abstract}

\section{Introduction}
\label{sec:intro}
Assessing the uncertainty and reliability of a  point or an event  of interest is an important but challenging task in many statistical and machine learning applications. Traditional approaches  often assume a specific distribution, or rely on asymptotic theory that requires a large sample size relative to the problem dimension, which,   in the big-data era, may not meet the challenges of  high dimensionality  or be  too rigid to accommodate various data imperfections. We would like to make the  inference \textit{data-based}   and \textit{method-driven} so that it can apply to any dataset and any estimator.   Notably, the method here may refer to an optimization criterion, a set of estimating equations, or a convergent algorithm. It turns out that the concept of data depth offers a universal nonasymptotic tool for robust estimation and inference  without having to specify a parametric density.


In 1975, John W. Tukey initiated the idea of location depth (or half-space depth) and demonstrated
its use in ranking multivariate data \citep{tukey1975mathematics}. Since then, a rich body of literature
on depth-based statistical methods  has emerged. Though conceptually simple, the powerful idea   extends to regression and more general setups
  \citep{Rousseeuw1999regression,ZHANG2002,mizera2002,Mizera2004,Muller2005,zuo2021general}.
In particular, \cite{ZHANG2002} studied a  general  class of score-function-based location depth and dispersion depth,  and   \cite{mizera2002} pointed out that half-space depth can be
criterion-driven, and proposed an operational tangent depth framework
when the criterion is differentiable. There also exist many other definitions
of data depth, simplicial depth
\citep{liu1990}, angular Tukey's depth \citep{Regina1992}, zonoid
depth \citep{koshevoy1997}, spatial depth \citep{Vardi2000} and
projection depth \citep{zuo2003}, to name a few. 
Data depth provides useful tools in quality
control \citep{Liu2003}, hypothesis testing \citep{Yeh1997,liu1999,Li2004},
outlier detection \citep{Becker1999},  data visualization \citep{Roussetal1999bagplot,Buttarazzi18}  and classification \citep{Jun2012,Lange2014,paindaveine2015,Subhajit2016}.
Despite the nice theoretical properties \citep{NOLAN1992,he1997,NOLAN1999,bai1999,zuo2000,chen2018robust,gao2020robust},  Tukey-type  depths suffer some serious issues that hinder their usage in real-life multivariate data.

Perhaps the biggest challenge lies in computation.
  \cite{JOHNSON1978} showed
that computing a given point's location depth is equivalent to solving
the closed hemisphere problem, thereby   NP-hard. Numerous  methods have been developed to compute   the exact depth    {}  in  low  dimensions   \citep{RUTS1996,Rousseeuw1998,ALOUPIS2002,Miller2003} and they are   mainly based on  enumeration or search. \cite{Liu2014} and \cite{DYCKERHOFF2016}
proposed more general algorithms with time complexity $\mathcal{O}(n^{m-1}\log n)$,
where $n$ is the sample size and $m$ is the dimensionality. Similarly, multivariate-quantile-based algorithms,  \cite{hallin2010}, \cite{Kong2012}, \cite{PAINDAVEINE2012}, have   algorithmic   complexity   exponentially large in $m$. The computation of  an estimate of  maximum depth  is even more challenging, and interested readers may  refer to \cite{Rousseeuw1998b}, \cite{Langerman2003a}, \cite{Langerman2003b} and \cite{Chan2004} among others.
In higher dimensions,  the class of   {approximate} methods  are more affordable and attractive   \citep{Rousseeuw1998,Dyckerhoff2004,Afshani2009,CHEN2013}.
They   often perform random sampling and   projection to reduce   the problem         to a lower-dimensional one,
but the required number of random subsets or projections is still combinatorially
large.
In experience, even for problems in moderate dimensions, 
  existing packages may either have poor accuracy or incur prohibitive computational costs. We refer to \cite{zuo2019new} and \cite{shao2020computing} for some recent  developments.

Moreover, in recent years, researchers
have   realized some severe scope limitations of Tukey-type   depths. For example,
for  multimodal distributions
or those with nonconvex density contours, some definitions of local depth might be more helpful; see  \cite{AGOSTINELLI2011}
and \cite{Davy2013}. Furthermore, modern optimization problems are often defined in a \textit{restricted} parameter space which may be curved, possess a low intrinsic dimension, or even  contain boundaries. Another important class of problems  emerging  from high-dimensional statistics have   \textit{nondifferentiable} objectives due to the use of  regularizers. Examples include variable selection,    low-rank matrix estimation,  and so on. In such contexts, how to introduce  data depth    is nontrivial, and has not been systematically studied before in the literature. 

This work investigates and extends Tukey's   depth from a subspace
learning viewpoint to overcome the aforementioned issues. We aim at   \textbf{operational} data depths with   efficient computation in multi-dimensions to advance the practice,  and hence,  abstract concepts   for pure theoretical purposes  are   not the focus.
Our main contributions are threefold. (i) A general framework of   problem-driven polished subspace depth, which emphasizes the roles of the underlying influence space and   discrepancy  measure,  is presented.      (ii) A  new matrix formulation enables us to utilize state-of-the-art optimization techniques including majorization-minimization,    iterative Procrustes rotations, and Nesterov's momentum-based acceleration to develop efficient algorithms for   depth computation  with   guaranteed fast convergence. (iii) Two approaches based on manifolds and  slack variables     extend  the notion    of depth significantly   to accommodate restricted parameter spaces and non-smooth objectives in possibly high dimensions.

In the first part of the work,   Section \ref{sec: Polished  depth} introduces the ``\textbf{Tukeyfication}'' process in detail and shows how Tukey's idea can be extended to define influence-driven  polished subspace depth. We also study its invariance  and give some illustrative   examples.  
Section \ref{sec:Data-depth-computation} studies optimization-based depth computation that scales up with problem dimensions and enjoys  a sound convergence guarantee.
Section \ref{sec:Simulation-Studies} performs extensive computer experiments. 
Some technical details and algorithmic details are left to the appendices.
The second part of the work is presented in our companion paper \citep{SheDepthII2021}, which investigates further extensions     via manifolds and slack variables to more sophisticated problems.
\paragraph{Notation. }
We use bold  symbols to denote vectors and matrices. A matrix $\boldsymbol{X}\in\mathbb{R}^{n\times p}$
is frequently partitioned into rows $\boldsymbol{X}=[\boldsymbol{x}_{1}\ldots\boldsymbol{x}_{n}]^{T}$
with $\boldsymbol{x}_{i}\in\mathbb{R}^{p}$. The vectorization   of   $\boldsymbol{X}$ is denoted
by $\textrm{vec}(\boldsymbol{X})\in\mathbb{R}^{np}$.
Let $\mathbb R_+=[0, +\infty]$.  Given   $\boldsymbol{X}\in\mathbb{R}^{n\times p}$, $\lVert\boldsymbol{X}\rVert_{F}$
and $\lVert\boldsymbol{X}\rVert_{2}$ denote its Frobenius norm and
spectral norm, respectively, $\| \bsbX\|_{\max}\triangleq  \max_{1\le i\le n, 1\le j\le p} |x_{ij}|$, and $\mbox{rank}(\bsbX)$ denotes its rank. The Moore-Penrose inverse of $\bsbX$ is denoted by $\bsbX^+$. The inner product of two matrices $\boldsymbol{X}$
and $\boldsymbol{Y}$ (of the same size) is defined as $\langle\boldsymbol{X},\boldsymbol{Y}\rangle=\textrm{Tr}(\boldsymbol{X}^{T}\boldsymbol{Y})$
and their element-wise product (Hadamard product) is $\boldsymbol{X}\circ\boldsymbol{Y}$.
  The Kronecker product is denoted by $\boldsymbol{X}\otimes\boldsymbol{Y}$
(where $\boldsymbol{X}$ and $\boldsymbol{Y}$ need not have the same
dimensions). Given  a set $\mathcal A \subset \mathbb R^{p\times m}$ and a matrix $\bsbT \in \mathbb R^{n\times p}$,    $\bsbT \circ \mathcal A = \{\bsbT \bsbA: \bsbA\in \mathcal A \}$. We use $\mathbb{O}^{m\times r}$ to represent  the set of all $m\times r$ matrices $\bsbV$ satisfying the orthogonality constraint $\bsbV^T\bsbV =\bsbI$.   For a vector $\bsb{a}=[a_1,\ldots,a_n]^{T}$ $\in \mathbb{R}^{n}$, $\mbox{diag}\{\bsb{a}\}$ is defined as an $n \times n$ diagonal matrix with diagonal entries given by $a_1,\ldots,a_n$, and for a square matrix $\bsb{A}=[a_{ij}]_{n\times n}$, $\mbox{diag}(\bsb{A}):=\mbox{diag}\{ a_{11},\ldots,a_{nn}\}$.  The indicator
function $1_{\mathcal{A}}(t)$ means $1_{\mathcal{A}}(t)=1$ if $t\in\mathcal{A}$ and
$0$ otherwise.
Given $f:\mathbb{R}^{n\times p}\rightarrow\mathbb{R}$,
  $f\in \mathcal C^1$ means that its Euclidean gradient  $\nabla f(\boldsymbol{X})$, an $n\times p$ matrix with the $(i,j)$ element
$ {\partial f }/{\partial x_{ij}}$,   exists and is continuous for any $\bsbX \in \mathbb R^{n\times p}$. Given two vectors $\bsb{\alpha}, \bsbb\in \mathbb R^p$, $\bsb{\alpha}\succeq \bsbb$ means $\alpha_j \ge \beta_j, 1\le  j \le p$ and $\bsb{\alpha}\succ \bsbb$ means $\alpha_j > \beta_j, 1\le  j \le p$.
\section{Half-space Depth and Tukeyfication}
\label{sec:Tukeyfication}
This section reviews half-space depth and extends it to polished subspace depth, which    comprises     three key elements:  influence function, influence space constraint, and discrepancy measure.

\label{sec: Polished  depth}
\subsection{Three elements for the polished half-space depth}

We begin with  a close examination of half-space depth.
Given $n$ observations $\boldsymbol{z}_{i}\in\mathbb{R}^{m}$, and  $\boldsymbol{\mu}^{\circ}$, a location of interest,
Tukey's
location   or empirical {half-space} depth   is the minimum number of sample points   enclosed by
a half-space containing $\boldsymbol{\mu}^{\circ}$:
$
d(\boldsymbol{\mu}^{\circ})=\min_{H\in\mathcal{H}(\boldsymbol{\mu}^{\circ})}{  \#}\{i:\boldsymbol{z}_{i}\in H\},
$
where $\mathcal{H}(\boldsymbol{\mu}^{\circ})$ is the set
of  all (closed) half-spaces  that cover $\boldsymbol{\mu}^{\circ}$.
(The conventional  definition  refers to
$d(\boldsymbol{\mu}^{\circ})/n$, but since we     study  data depth  associated with $n$ observations,
 all  trivial multiplicative factors and additive
constants are  dropped for simplicity unless otherwise specified.)
Motivated by Section \ref{sec:intro}, a pressing question
is     to extend this nonparametric tool to any given estimation method.

Below, we work in a  {\textbf{supervised}} setup  with     $n$ (approximately) i.i.d.  observations of  $m$ response variables  and $p$ predictor variables  $(\bsby_i, \bsbx_i  )\in{  \mathcal S} \subset  \mathbb R^m\times \mathbb R^p$ ($1\le i \le n$), and     $\mathcal S$ is referred to as the ambient \textit{sample space}. In the  special case of $m$-dimensional  location estimation, where  there are only observations $\bsby_i$  available but no nontrivial predictor variables (i.e.,    $x_i=1$,   $1\le i \le n$), the   sample space is characterized by   $\bsby_i\in  \mathcal S \subset \mathbb R^m$ by convention.

Let  $\boldsymbol{X}=[\boldsymbol{x}_{1}\ldots\boldsymbol{x}_{n}]^{T}\in\mathbb{R}^{n\times p}$,  $\boldsymbol{Y}=[\boldsymbol{y}_{1}\ldots\boldsymbol{y}_{n}]^{T}\in\mathbb{R}^{n\times m}$, and  $\boldsymbol{B}$ be the unknown parameter matrix to estimate. Suppose that the estimation method is specified  by a set of estimating equations:
\begin{equation}
\sum_{i=1}^{n}\boldsymbol{T}(  {\boldsymbol{B}};\boldsymbol{x}_{i},\boldsymbol{y}_{i})=\boldsymbol{0}.  \label{eq:sample addtivity}
\end{equation}
Eqn. \eqref{eq:sample addtivity} can be derived from an optimization problem $\min_{\boldsymbol{B}}f(\boldsymbol{B};\boldsymbol{X},\boldsymbol{Y})$, which is often our starting point in this paper. For example, assuming
\begin{align}
f (\boldsymbol{B};\boldsymbol{X},\boldsymbol{Y})=\sum_{i}l(\boldsymbol{B};\boldsymbol{x}_{i},\boldsymbol{y}_{i}), \label{eq:addloss}
\end{align} with the same loss $l\in \mathcal C^1$ (which need not  be a negative likelihood function) applied and summed on $n$ approximately i.i.d. sample points, we get    $\boldsymbol{T}(\boldsymbol{B};\boldsymbol{x}_{i},\boldsymbol{y}_{i})=\nabla_{\boldsymbol{B}}l(\boldsymbol{B};\boldsymbol{x}_{i},\boldsymbol{y}_{i})$. 
 However, in the presence  of a  regularizer added in the criterion,  the associated estimation equations may not always have the pleasant   sample-additive form  \citep{SheDepthII2021}.

As pointed out by Peter Rousseeuw and anonymous reviewers, in the above setup, $\boldsymbol{T}(\cdot)$ is proportional to the influence function \citep{hampel2011robust}, and so we call  $ \boldsymbol{T}(\boldsymbol{B};\boldsymbol{x}_{i},\boldsymbol{y}_{i})$ (or ${\bsb{T}}_i(\bsbB)$, for short) the influence at observation $i$. We  further assume that
$
{\bsb{T}}_i(\bsbB)$ is in an \emph{influence space} ${ \mathcal G}\subset   \mathbb R^{p\times m}
$.
Of course,  in many applications  one can directly define the influences or estimating equations
without involving  an explicit   objective, sometimes from   an iterative algorithm or a surrogate function.

Let  $\boldsymbol{B}^{\circ}$ be any given point
in the \textit{parameter space} $\Omega\subset \mathbb R^{p\times m}$ and   $\boldsymbol{T}_{i}^{\circ}=\boldsymbol{T}(\boldsymbol{B}^{\circ};\boldsymbol{x}_{i},\boldsymbol{y}_{i})$. Mimicking Tukey's location depth, we first
project the influences onto a line  with direction $\bsbV$, and then measure how the estimating equations are maintained   via a discrepancy
function $\varphi$. This results in
the following \textit{polished half-space depth} (\textbf{PHD})  \begin{equation}
\mbox{\textbf{PHD:}} \quad d_\varphi(\boldsymbol{B}^{\circ})=\min_{\boldsymbol{V} } \sum_{i}\varphi(\langle\boldsymbol{V},\boldsymbol{T}_{i}^{\circ}\rangle)\;\textrm{s.t.}\;\lVert\boldsymbol{V}\rVert_{F}=1, \boldsymbol{V}\in \bar{\mathcal G},
\label{eq:PhD}
\end{equation}
where $\bsbV$ is restricted in   a projection space $\bar{\mathcal G}$.   We call \eqref{eq:PhD} ``polished'', owing to (i) the flexibility of $\varphi$,
which need not be a monotone function in particular, and (ii) the additional requirement  $\boldsymbol{V}\in\bar{\mathcal G}$, to complete the notion of depth   necessary for defining, for example, covariance depth and Riemannian manifold depth. Although  $\bar{\mathcal G}$  can be much more general, we set
$ \bar {\mathcal G} = \mathcal G$ throughout the work, and the corresponding  \textit{influence space constraint}
$\boldsymbol{V}\in\mathcal G$ is perhaps natural seen from the inner product $\langle\boldsymbol{V},\boldsymbol{T}_{i}^{\circ}\rangle$. 
     We occasionally write   $d_\varphi(\boldsymbol{B}^{\circ}; \{\boldsymbol{T}_{i}^{\circ}\}, \mathcal G)$ to emphasize its dependence on $\{\boldsymbol{T}_{i}^{\circ}\}$ and $ \mathcal G$.
For more discussions of the inner product, projection, and  constraint,   see Section \ref{subsec:RieDep} of \cite{SheDepthII2021} for a  general   ``{directional directive}'' or  {``geodesic''} framework. For supervised problems, a    trace form amenable to matrix optimization will be introduced in Section \ref{sec:Data-depth-computation}.  Also, the   criterion  in \eqref{eq:PhD} can  be extended to a   U-statistic form.

 Special case: when $\varphi(t) = 1_{\ge 0} (t)$, we abbreviate $d_{\varphi}$ as $d_{01}$. (Although $1_{\ge 0}$ is conventionally used, $  0.5 \cdot 1_{=0} +     1_{>0}  $  is perhaps a better choice for defining $d_{01}$ \citep{SheDepthII2021}, and is more convenient in the successive optimization in Section \ref{sec:Data-depth-computation}.) Consider a Gaussian location estimation problem that defines the loss of the unknown location $\bsbmu\in \Omega=\mathbb R^m$  as  $l(\boldsymbol{\mu} ;\boldsymbol{z}_{i})=\|\boldsymbol{\mu}-\boldsymbol{z}_{i}\|_2^{2}/2$,  for  $n$   observations $\bsbz_i\in \mathcal S= \mathbb R^{m}$, then,   $\boldsymbol{T}(\boldsymbol{\mu}^{\circ}; \boldsymbol{z}_{i})=\nabla l(\bsbmu; \bsbz_i)|_{\bsbmu = \bsbmu^\circ}=\boldsymbol{\mu}^{\circ}-\boldsymbol{z}_{i} \in \mathcal G = \mathbb R^m$, and so $d_{01}$ based on     \eqref{eq:PhD}  becomes  Tukey's  location depth. Similarly, for the ordinary single-response regression, where  $m=1$ and the loss is quadratic:  $l(\boldsymbol{\beta} ;\boldsymbol{x}_{i},y_{i})=(\boldsymbol{x}_{i}^{T}\boldsymbol{\beta} -y_{i})^{2}/2$, simple calculation shows   $\boldsymbol{T}(\boldsymbol{\beta}^{\circ};\boldsymbol{x}_{i}, {y}_{i})= (\boldsymbol{x}_{i}^T \boldsymbol{\beta}^{\circ  }- {y}_{i})\boldsymbol{x}_{i} $,  corresponding to  the celebrated regression depth \citep{Rousseeuw1999regression}. The sample additive form of the objective in \eqref{eq:PhD} makes it possible to define a population version with respect to a certain distribution  $F$, in place of the empirical distribution, but we focus on the sample version without assuming a distribution for the data or an infinite sample size.

The three essential elements in defining  \eqref{eq:PhD}, namely, $\bsbT_i^\circ$, $\varphi$, and $\mathcal G$,  deserve a more careful discussion. The influence at observation $i$, not always taking the plain difference  $\boldsymbol{\mu}^{\circ}-\boldsymbol{z}_{i}$ as in location depth, can be derived from   any     criterion. So the influences may be  rooted in a parametric model (such as a Gaussian one), but Tukey's   mechanism, which we will refer to as ``\textbf{Tukeyfication}'', offers    nonparametricness   and  robustness.
In this sense,  
    \eqref{eq:PhD}  shares similarities with Owen's empirical likelihood \citep{owen2001} which also    operates on  a given set of     estimating equations for   nonparametric inference, but   can be  more robust---for instance,  $d_{01}$  targets ``Tukey's  median'' (far more robust than the $\ell_1$-median), instead of the ``mean'' under \eqref{eq:sample addtivity} or  maximum likelihood estimation. However, when the problem under consideration has   nondifferentiability or additional constraints, which is common in high-dimensional statistics and machine learning, the influences  must  be adjusted, which will be examined in our companion paper \citep{SheDepthII2021}.

The influence space $\mathcal G$ is often a linear subspace. Under \eqref{eq:addloss}, when $\mathcal G$ is trivially $   \mathbb R^{p\times m}$ and $l$ is differentiable, the influence space constraint  in \eqref{eq:PhD} is   inactive and   $d_{01}$   is    in the   framework of \textit{tangent
depth} \citep{mizera2002}.  In general, however, the role of     $\mathcal{G}$ cannot  be     ignored especially in some matrix problems,     covariance estimation,   multivariate meta analysis  and   manifold-restricted learning, among others, which gives an important distinction     from      many depth definitions. We feel that it is necessary to differentiate the sample space, parameter space, and influence space in studying the  concept of data depth. The three spaces need not be identical, although      for Tukey's location depth, $\mathcal S = \Omega = \mathcal G = \mathbb R^{ m}$. But when $\mathcal G$ is not simply the full Euclidean space,      one may want to impose some more {structural} properties on $\bsbV$.

 With regards to  the necessity and benefit of introducing $\varphi$,   we notice that the $0$-$1$   loss, though   scale free,  penalizes
  projected influences with a constant cost and
 thus suffers some issues. Specifically, it is non-smooth, the magnitude information of the
influences is not taken into  account, and the   dichotomous    measurement
may be crude and unstable for influences near  zero.    To see what other forms  $\varphi$ can take,  let us assume $\mathcal G=\mathbb R^{p\times m}$ and rewrite the original half-space depth    $d_{01}$    to gain more insights:    
 \begin{align}
d_{01}(\boldsymbol{B}^{\circ})
   =\min_{\lVert\boldsymbol{V}\rVert_{F}=1}\sum_{}1_{\ge 0}(\langle\boldsymbol{V},\boldsymbol{T}_{i}^{\circ}\rangle)
   =\min_{\lVert\boldsymbol{V}\rVert_{F}=1}\sum_{}1_{\leq0}(\langle\boldsymbol{V},\boldsymbol{T}_{i}^{\circ}\rangle).
    \label{eq:half-space depth} \end{align}
      The latter form     studies   a binary classification problem      on the   margins $\langle\boldsymbol{V},\boldsymbol{T}_{i}^{\circ}\rangle$;
other classification losses, such as  the hinge loss,  logistic deviance, and   the Savage loss    \citep{Hamed2009} can be possibly used.
The classification viewpoint enables us to borrow some tools in machine learning for nonasymptotic theoretical analysis. Also, seen from the first expression, one can   replace    the degenerate $1_{\ge0}(t) $   for a point mass at zero by   any distribution function, and choosing a continuous one  can bring in some smoothing effect.

Another useful $\varphi$-family is from   the  ``$\psi$-functions'' in M-estimation. (In fact, assuming    $\boldsymbol{T}(\boldsymbol{\mu}^{\circ}; \boldsymbol{z}_{i})=\boldsymbol{\mu}^{\circ}-\boldsymbol{z}_{i} $ in the location setup, $d_\psi$ defined in \eqref{eq:PhD} is the unscaled   generalized Tukey depth due to  \cite{ZHANG2002}; see \eqref{psd-rC} for our new proposal for   handling the scale issue.)  Our motivation is from the ``contrast'' representation of \eqref{eq:half-space depth}
\begin{align}
d_{01}(\boldsymbol{B}^{\circ})   = ({n}/{2})+ ({1}/{2})\min_{\lVert\boldsymbol{V}\rVert_{F}=1, \bsbV\in \mathcal G}\sum_{i}\sgn(\langle\boldsymbol{V},\boldsymbol{T}_{i}^{\circ}\rangle)
, \label{eq:PHD_contrast}
\end{align}
where       $    \sgn(t)\triangleq1_{\geq0}(t)-1_{<0}(t)$
is just the $\psi$-function associated with  the $\ell_{1}$-norm loss except that   $\sgn(0)=1$.  \cite{ZHANG2002} studied some theoretical properties when using a monotone $\psi$  (such as Huber's $\psi$). Interestingly, it seems that        \textit{redescending} $\psi$-functions that are non-monotone
\citep{hampel2011robust}, and their \emph{rectified} versions  $\max\{0, \psi(t)\}$ in particular,  are potentially useful in    dealing with data that are not unimodal; see Figure \ref{1D local depths}  in Section \ref{subsec:conesubspaceinv}.
\subsection{Polished subspace depth and invariance}
\label{subsec:conesubspaceinv}
The ideas of    projection and polishing apply more generally.     For example, we can extend Tukey's   straight line projection   to a subspace projection to improve outlier resistance. Toward this, introduce \textit{vectorized} influences   \begin{align}\bsbt_i^\circ = \vect (\boldsymbol{T}_{i}^{\circ}),\end{align} and  assume they  are in some influence space denoted by $  \mathcal G$,  a subset of  $  \mathbb R^{pm}$, with a slight abuse of notation. Using      $K$,              a {proper cone}  \citep[page 43]{Boyd2004}   that induces a partial ordering on    $\mathbb R^r$ ($r\le pm$)  to sort the projected influences, we can define a \textit{projected cone depth}     by
\begin{align}
\min_{\boldsymbol{V} = [\bsbv_1, \ldots, \bsbv_r]\in \mathbb R^{pm\times r}}\sum_{i}1_{K} ( \boldsymbol{V}^T \boldsymbol{t}_{i}^{\circ} ) \mbox{ s.t. }  \bsbV^T \bsbV = \bsbI_{r\times r},   \bsbv_s\in\mathcal G, 1\le s \le r.\end{align}

In the particular case of $K = \mathbb R_+^r$, a smooth ${\varphi}: \mathbb R^r \rightarrow \mathbb R$     in place of $1_K$ gives the \textit{polished subspace depth} (\textbf{PSD})
which includes the {polished half-space depth}  \eqref{eq:PhD} as $r=1$:
 \begin{align}
\mbox{\textbf{PSD:}}  \quad d_{ \varphi, r  } (\bsbB^\circ) \  = \quad \begin{split}
 & \min_{\boldsymbol{V} = [\bsbv_1, \ldots, \bsbv_r]\in \mathbb R^{pm\times r}} \sum_{i} \prod_{s=1}^r \varphi  ( \boldsymbol{v}_s^T \boldsymbol{t}_{i}^{\circ} ) \\ & \qquad\  \ \mbox{ s.t. }  \bsbV^T \bsbV = \bsbI_{r\times r},   \bsbv_s\in\mathcal G, 1\le s \le r.
 \end{split} \label{psd-r}
\end{align}
When necessary, we also write the depth as  $d_{\varphi, r} (\bsbB^\circ; \{ \bsbt_i^\circ\}, \mathcal G)$. It is easy to prove that $d_{01, r}$ is non-increasing in $r$. 

To measure the errors more  precisely, it is necessary to violate  $\varphi(t/\sigma) = \varphi(t)$  $\forall \sigma>0$ (for it would mean that when $r=1$,     $\varphi(t)$   must be constant as  $t>0$ or $t<0$, i.e.,  a sign-type function though not necessarily symmetric). Then how do we  achieve scale invariance?
   \cite{ZHANG2002} proposed a scaled form to maintain invariance for location depth, where  $\mathcal G$ is the full Euclidean space and    $\sigma(\cdot)$ is  a scale-equivariant function
\begin{equation}
 \min_{\boldsymbol{v} } \sum_{i}\varphi\left( \frac{\boldsymbol{v}^T  (\boldsymbol{\mu}^{\circ} - \bsbz_i)}{\sigma(\{\boldsymbol{v}^T  (\boldsymbol{\mu}^{\circ} - \bsbz_i)\}_{i=1}^n)}\right)\;\textrm{s.t.}\;\lVert\boldsymbol{v}\rVert_{2}=1.
\label{univscaledform-zhang}
\end{equation}
     But \eqref{univscaledform-zhang} is barely operational   from an optimization perspective, because     $\bsbv$ is involved inside $\sigma$, while    $\sigma$ may be nonsmooth or even lack an explicit formula. Moreover, how  to extend \eqref{univscaledform-zhang}  to $r>1$ is unclear.
We give a   simple but effective modification of  \eqref{psd-r} as follows.

First, our goal is  to study the  invariance   of a  {general} $\varphi$-depth     under some \textit{transformations of the (vectorized) {influences}}. For example, it is preferable  to maintain the  depth value when switching to  {scaled} influences $\bsbt_i^\circ\rightarrow k \bsbt_i^\circ$ for all $k\in \mathbb R$, or even {affine-transformed} influences $\bsbt_i^\circ \rightarrow \bsbA\bsbt_i^\circ$ for all nonsingular  $\bsbA \in \mathbb R^{pm\times pm}$.  For some related invariance studies in the scenarios  of location  depth and regression depth,   refer to   \cite{zuo2000} and  \cite{zuo2021general}. 

Let \begin{align}\bar \bsbT^\circ = [\bsbt_1^\circ, \ldots, \bsbt_n^\circ]^T\in \mathbb R^{n\times pm}\label{vecMatT}
\end{align}
be the matrix formed by vectorized influences.
We observe that  $d_{\varphi, r}$  defined in \eqref{psd-r}  does enjoy some sort of  orthogonal invariance:       for \textit{all} $\varphi$, $r$, and $ \mathcal G$,        \begin{align}
d_{\varphi, r} (\bsbB^\circ; \{\bsbA  \bsbt_i^\circ\}, \bsbA \circ \mathcal G) = d_{\varphi, r} (\bsbB^\circ; \{ \bsbt_i^\circ\}, \mathcal G) ,  \quad \forall \bsbA \in \mathbb O^{pm\times pm}.  \label{orthinv}\end{align}
 In fact, $d_{\varphi, r} (\bsbB^\circ; \{ \bsbt_i^\circ\}, \mathcal G) $ can be defined as $\min  \langle \bsb1, \varphi(\bar\bsbT^\circ \bsbV)\rangle \mbox{ s.t. } \bsbv_s \in \mathcal G,  \bsbV ^T \bsbV \allowbreak = \bsbI $ with
  $\varphi(\bar\bsbT^\circ \bsbV) = [\varphi(\bsbV^T\bsbt_1^\circ), \ldots, \varphi(\bsbV^T\bsbt_n^\circ)]^T$; substituting    $\bar\bsbT^\circ \bsbA^T$ for $\bar\bsbT^\circ$ and $\bsbA \bsbV$ for $\bsbV$ keeps the problem unchanged.

  Motivated by \eqref{orthinv}, we define an  {\emph{invariant}} form of polished subspace depth
\begin{align}
d_{ \varphi, r  }^{\bsbC}(\bsbB^\circ) =  \min_{\boldsymbol{V} \in \mathbb R^{pm\times r}}\sum_{i} {\varphi} ( \boldsymbol{V}^T \boldsymbol{t}_{i}^{\circ} ) \mbox{ s.t. }  \bsbV^T \bsbC(\bar\bsbT^\circ)\bsbV = \bsbI_{r\times r},   \bsbv_s\in\mathcal G, 1\le s \le r, \label{psd-rC}
\end{align}
where  $\bsbC(\bar\bsbT^\circ)$ is positive semi-definite  and affine equivariant in the sense that \begin{align}\bsbC(\bar\bsbT^\circ  \bsbA^T) = \bsbA\bsbC (\bar\bsbT^\circ ) \bsbA^T\label{CAffinv}\end{align} for any nonsingular $\bsbA \in \mathbb R^{pm\times pm}$, and $\mbox{rank}(\bsbC(\bar\bsbT^\circ))\ge r$.   Then it can be easily shown  that   for {any} $\varphi$, $r$, $\mathcal G$,
\begin{align}
d_{\varphi,r }^{\bsbC}(\bsbB^\circ; \{\bsbA  \bsbt_i^\circ\}, \bsbA \circ \mathcal G) = d_{\varphi, r}^{\bsbC}(\bsbB^\circ; \{ \bsbt_i^\circ\}, (\bsbA^T \bsbA) \circ\mathcal G)    \label{affinv}
\end{align}
for all nonsingular $ \bsbA \in \mathbb R^{pm\times pm}$.
Therefore, if $\mathcal G$ is a cone satisfying $a \mathcal G = \mathcal G, \forall a>0$, $d_{\varphi,r }^{\bsbC}$ has the desired  scale invariance: $d_{\varphi,r }^{\bsbC}(\bsbB^\circ; \{k  \bsbt_i^\circ\}, k \mathcal G) =  d_{\varphi, r}^{\bsbC}(\bsbB^\circ; \allowbreak\{ \bsbt_i^\circ\}, \mathcal G)$ for any $k\in \mathbb R$. Moreover, when $\mathcal G$ is the full Euclidean space, like in location depth or regression depth, $(\bsbA^T \bsbA) \circ\mathcal G=\mathcal G$ holds for all nonsingular $ \bsbA \in \mathbb R^{pm\times pm}$, and so          $d_{\varphi,r }^{\bsbC}$ is affine invariant,   as \eqref{univscaledform-zhang}, but for all $r$.

Another appealing fact of \eqref{psd-rC} is that compared with  the basic form $d_{\varphi, r}$ (cf.    \eqref{psd-r}),  it adds little  cost in computation. When  $\mathcal G$ is  Euclidean, one can convert  $d_{\varphi, r}^{\bsbC}$  to  $d_{\varphi, r}$   with a reparametrization    $\bsbV' = \bsbD^\circ \bsbV^{\circ T} \bsbV$, where $\bsbD^\circ, \bsbU^\circ$ are obtained from the spectral decomposition $\bsbC(\bar\bsbT^\circ) = \bsbV^\circ(\bsbD^{\circ  })^2 \bsbV^{\circ T}$.
Specifically,   we can simply define  $d_{\varphi, r}$   on the column space basis    $\bsbU^\circ$ of  $\bar\bsbT^\circ$ (consisting of all     left singular vectors of the matrix of vectorized influences), which amounts to $d_{\varphi, r}^{\bsbC}$ using    $\bsbC(\bar\bsbT^\circ) =  (\bar\bsbT^\circ)^T\bsbT^\circ$  that obviously satisfies  \eqref{CAffinv}. Based on the previous discussion, this \textit{normalized} version   has affine invariance regardless of  $\varphi$  in use.   Moreover,  owing to the orthogonal invariance, we can prove    that the optimization problem depends on $\bsbU^\circ$ through $\bsbU^\circ \bsbU^{\circ T}$, and so the choice of $\bsbU^\circ $   will not affect our depth.

\begin{figure}[htp!]
\centering
\subfloat {\includegraphics[width=.9\columnwidth]{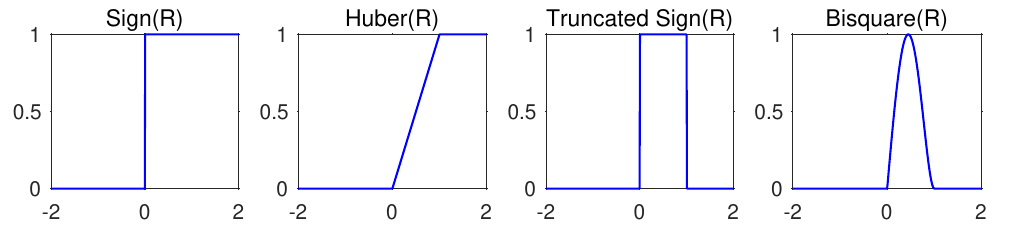}} \\
\setcounter{subfigure}{0}
\subfloat[\scriptsize  Examples of one-sided  $\varphi$ functions (top row) and the  corresponding depth values (bottom row, with a factor of $1/n$).  Tukey's depth uses the 0-1 loss or the rectified
sgn function (1st column).  The depth curve with rectified Huber (2nd column) is a smoothed version of it. In the 3rd column, the rectified truncated sign function, which is non-monotone, is used  as $\varphi$  to generate a local depth curve.
In the last  column, with Tukey's bisquare function rectified, a similar local depth curve is obtained with the dashed lines labeling some deepest points. \label{1D local depths one side}  ]
{\includegraphics[width=.9\columnwidth]{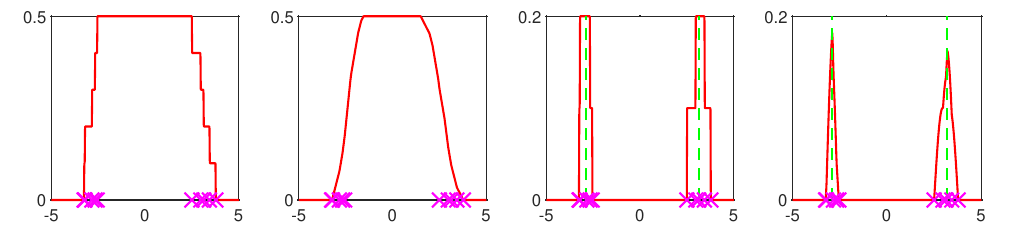}}

\subfloat{\includegraphics[width=.9\columnwidth]{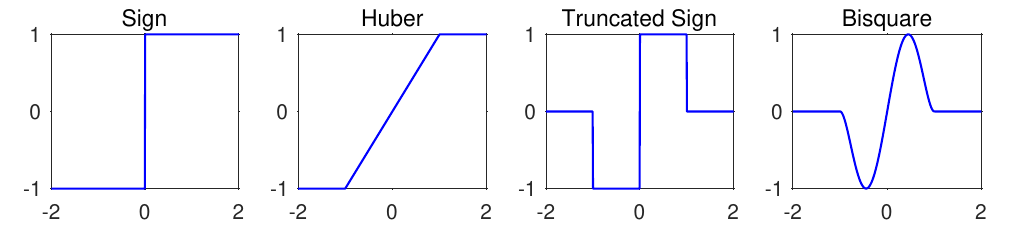}} \\
\setcounter{subfigure}{1}
\subfloat[\scriptsize Examples of two-sided $\varphi$ functions (top row)  and the  corresponding depth values (bottom row, with a factor of $1/n$). 1st column: The sign function leads to the same Tukey's depth as the one-sided sign.   2nd column: Huber's $\psi$ smoothes Tukey's depth, but behaves differently from  rectified Huber in (a),  say at the points lying outside the support of  the data. In the last two columns, redescending functions (without rectification) are used, and some  {shallowest} points that resemble the cluster boundaries are labeled with  dashed lines. \label{1D local depths two side} ]
{\includegraphics[width=.9\columnwidth]{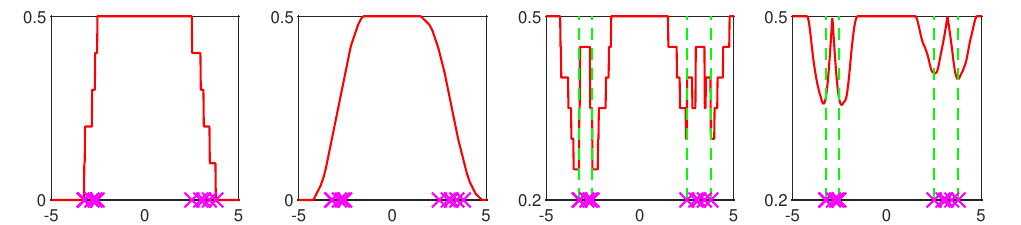} }
\caption{\small An illustration  of some $\varphi$ functions (one sided  and two sided) and their corresponding
depth values on a one-dimensional    dataset with the data points  denoted by crosses. 
 \label{1D local depths}}
\end{figure}

Finally,  we illustrate the role of  $\varphi$ in  revealing different  characteristics of a dataset with
Figure \ref{1D local depths}. The data points, denoted by crosses, are generated according to a Gaussian mixture model,   $y_{i}\stackrel{}{\sim}0.5\mathcal{N}(-3,1/16)+0.5\mathcal{N}(3,1/4), 1\le i \le 10$.  We tried   some ``two-sided'' $\varphi$'s in the contrast form \eqref{eq:PHD_contrast},  constructed from the following $\psi$-functions  widely adopted in robust statistics \citep{hampel2011robust}: the sign  $\psi(t) = \sgn(t)$ (note however that $\sgn(0)=1$),   Huber's $\psi(t) = t 1_{|t|\le c} + c \sgn(t) 1_{|t|>c}$, the truncated sign $\psi(t)= \sgn(t) 1_{|t|\le c}$ and Tukey's bisquare $\psi(t) = t (1-(t/c)^2)^2 1_{|t|\le c} $, where we set $c=1$ and then scaled all of them to have a range $[-1,  1]$.
 We also tested some ``one-sided'' $\varphi$'s  in \eqref{eq:PhD}  defined via $\psi$:
  $\varphi(t)=\max\{0, \psi(t)\}$, which we  call   \textit{rectified} $\psi$'s. The rectified truncated sign
is  also considered  in \cite{AGOSTINELLI2011}, and is  called the \textit{slab} function.   
 The results for one-sided  $\varphi$'s are   shown
in Figure \ref{1D local depths one side})  and those for two-sided $\varphi$'s are in Figure   \ref{1D local depths two side}).

According to the figure, Tukey's depth can be achieved  using the sign or rectified sign and    smoothened by  a continuous $\varphi$ (like the ones via Huber's $\psi$). Moreover, the  {rectified  redescending} $\psi$'s
 offer some \textit{local}
depths on the bimodal dataset, which deserves further investigation.  
How to choose a proper  $\varphi$   to  discover desired  data features, and whether there is a universal recommendation with certain optimality    are beyond the scope of the paper, but we will see that introducing $\varphi$-depth greatly assists  computation.

\subsection{Examples}
\label{subsec:ex}
 In the following, we provide some instances
in different statistical contexts.


\paragraph{GLM depths}
Consider   a  \textit{vector} generalized linear model (GLM) with  a  cumulant
function $b $ and  the {canonical} link $g = (b')^{-1}$. Then $l(\boldsymbol{B};\boldsymbol{x}_{i},\boldsymbol{y}_{i})=-\langle\boldsymbol{B}^{T}\boldsymbol{x}_{i},\boldsymbol{y}_{i}\rangle+\langle\boldsymbol{1},b(\boldsymbol{B}^{T}\boldsymbol{x}_{i})\rangle$, where  $b$ is applied componentwise.
The estimation equations are given by
\begin{align}
\bsbX^T (b' (\bsbX \bsbB)  - \bsbY) = \bsb0, \label{eq:vecGLMEE}
\end{align}
and $ \bsbT_i(\bsbB )=\boldsymbol{x}_{i} (b^{\prime}(\boldsymbol{B}^{T}\boldsymbol{x}_{i})-\boldsymbol{y}_{i})^T \in \mathcal G = \mathbb R^{p\times m}$,
and so (\ref{eq:PhD}) becomes
\begin{align}
d_\varphi(\boldsymbol{B}^{\circ})   =&\min_{ \lVert\boldsymbol{V}\rVert_{F}=1}\sum_{i}\varphi(\boldsymbol{x}_{i}^{T}\boldsymbol{V}[b^{\prime}(\boldsymbol{B}^{\circ T}\boldsymbol{x}_{i})-\boldsymbol{y}_{i}]) \label{eq:polished regression data depth}
\end{align}
where $b^{\prime}(\cdot)$ and $\varphi(\cdot)$ are applied element-wise.

First, under the classical Gaussian assumption,     $b^{\prime}(\cdot)$ is the identity function, and so  (\ref{eq:polished regression data depth})
 covers  the multivariate regression depth \citep{Bern2002}. How to      incorporate   dependence   into   data depth, as raised by    \cite{Eddy1999}, is a meaningful  question. But  under $\bsby_i \sim \mathcal N(\bsbB^T \bsbx_i, \bsbSig)$, the weighted criterion  for estimating $\bsbB$ is  $\mbox{Tr}\{ (\bsbY - \bsbX \bsbB)\bsbSig^{-1} (\bsbY - \bsbX \bsbB)\}/2$, and thus \eqref{eq:vecGLMEE} becomes  $\bsbX^T (\bsbX \bsbB  - \bsbY) \bsbSigma^{-1} = \bsb0$ or $\sum \boldsymbol{x}_{i} ( \boldsymbol{B}^{T}\boldsymbol{x}_{i}-\boldsymbol{y}_{i})^T  \bsbSigma^{-1}  = \bsb0$. Therefore,   adopting an   affine invariant depth indicates  no need to take the between-response covariance  into consideration.

Next, let us consider non-Gaussian GLMs. When $m=1$   and $\varphi(t)=\sgn  (t)$, it is well known that the associated GLM depth
amounts to applying regression depth on the transformed observations $(g(y_{i}), \bsbx_i)$  owing to        the property:     $\sgn ( \boldsymbol{x}_{i}^T \boldsymbol{v} ( b^{\prime}(\boldsymbol{x}_{i}^{T}\boldsymbol{\beta})-y_{i}))= \sgn  (\langle\boldsymbol{v},\boldsymbol{x}_{i}\rangle)  \sgn (u(b^{\prime}(\boldsymbol{x}_{i}^{T}\boldsymbol{\beta}))-u(y_{i}))$
for any  strictly increasing $u$  \citep{Van2002}. However, we remark that the     monotone invariance property  does not hold in general for multivariate problems ($m>1$), and so GLM depths do have their value.   
We can  also use the logistic regression depth to illustrate the weakness of $\varphi(t)=\sgn  (t)$. Let $m=1$,     $r_{i}=b^{\prime}(\boldsymbol{x}_{i}^{T}\boldsymbol{\beta})-y_{i}=\exp(\boldsymbol{x}_{i}^{T}\boldsymbol{\beta})/(1+\exp(\boldsymbol{x}_{i}^{T}\boldsymbol{\beta})) - y_i$. For such binary $y_i$,  the sigmoidal $r_i$    appear more reasonable  than the difference-based residuals
$ \boldsymbol{x}_{i}^{T}\boldsymbol{\beta}  - y_i$   in regression. But  because          $\sgn (  r_{i}) = 1 - 2  y_i$ (regardless of the difference between $\exp(\boldsymbol{x}_{i}^{T}\boldsymbol{\beta})/(1+\exp(\boldsymbol{x}_{i}^{T}\boldsymbol{\beta})) $ and $ y_i$), the resulting depth does not vary with    $\bsbb\in \mathbb R^p$ as long as it is finite,  an  evidence of the crudeness of $d_{01}$ in this scenario.


 Finally, we point out that   although one could vectorize \eqref{eq:vecGLMEE} via $\bsby = \vect (\bsbY)\in \mathbb R^{nm}$,  $\bsbb = \vect (\bsbB)\in \mathbb R^{pm}$ and  $\bsbZ = \bsbI\otimes \bsbX$ to get   \begin{align}\bsbZ^T (b' (\bsbZ \bsbb)  - \bsby) = \bsb0,\end{align}  the associated data depth would not  have a valid population definition. In fact,        $\bsbZ$  has a   block diagonal form, meaning that  its rows   cannot  be treated as observations following  the same distribution, and  the vectorized equations do not have the desired sample additivity  on $(\bsbx_i, \bsby_i)$,  $1\le i \le n$.
Introducing  data depth via  the   generalized estimating equations (GEEs) \citep{Liang86} may suffer the same issue. Concretely,  the GEEs  for our problem are given by \begin{equation}\label{gee-general}
\begin{aligned}
 (\bsbI\otimes \bsbX)^T  & \mbox{diag}\{ (b'')^{1/2}(\vect (\bsbX \bsbB ))\}\ \times\
\bsbW^{-1} \times    \\&  \mbox{diag}\{ (b'')^{-1/2}(\vect (\bsbX \bsbB ))\}(b'(\vect(\bsbX \bsbB )) - \vect (\bsbY))=\bsb0,
 \end{aligned}\end{equation} where   the working correlation matrix $\bsbW=  \bsbSigma \otimes \bsbI$  with $\bsbSigma$ known (say, the    sample correlation matrix of $\bsbY$ or some regularized estimate). In the special case that  $b'$ is identity or  $\bsbSig$ is diagonal, $\{ (b'')^{1/2}(\vect (\bsbX \bsbB^\circ))\}$ and $\mbox{diag}\{ (b'')^{-1/2}(\vect (\bsbX \bsbB^\circ))\}$   cancel, and \eqref{gee-general} can be rephrased as $\bsbX^T (b'(\bsbX \bsbB)  - \bsbY) \bsbSigma^{-1} = \bsb0$, which \textit{is} sample additive. But the property  holds no longer for   multivariate  non-Gaussian  GEEs  to induce a legitimate data depth.

\paragraph{Covariance depth}

Assume that  $\textrm{vec}(\boldsymbol{Y})\sim\mathcal{N}(\boldsymbol{0},\boldsymbol{\Sigma}\otimes\boldsymbol{I})$ for $\boldsymbol{Y}\in\mathbb{R}^{n\times m}$,
where the between-column covariance matrix $\boldsymbol{\Sigma}$
is positive definite. Let   $\boldsymbol{W}=\boldsymbol{\Sigma}^{-1}$.
From the negative log-likelihood  $l(\boldsymbol{W};\boldsymbol{y}_{i})=(\textrm{Tr}(\boldsymbol{W}\boldsymbol{y}_{i}\boldsymbol{y}_{i}^{T})-\log\det\boldsymbol{W})/n$  (up to some scaling
and additive constants),  we know that its gradient  takes a simple difference form   $  (\boldsymbol{y}_{i}\boldsymbol{y}_{i}^{T}-\boldsymbol{\Sigma} )/n$, symmetric but not necessarily positive semi-definite.
 The depth for a positive definite   $\boldsymbol{\Sigma}^{\circ}$  
 according to  \eqref{eq:PhD} is
\[
d_{\varphi }(\bsbSig^\circ)=\min_{\boldsymbol{V}}\sum_{i}\varphi(\boldsymbol{y}_{i}^{T}\boldsymbol{V}\boldsymbol{y}_{i}-\langle\boldsymbol{V},\boldsymbol{\Sigma}^{\circ}\rangle), \mbox{ s.t. } \lVert\boldsymbol{V}\rVert_{F}=1,\boldsymbol{V}=\boldsymbol{V}^{T},
\]
where $\bsbV$   is additionally required to be symmetric, as an outcome
of the symmetry of the gradient.
Adding a further rank restriction:  $\textrm{rank}(\boldsymbol{V})=1$,  $  \boldsymbol{V}$  simplifies to $ \pm\boldsymbol{v}\boldsymbol{v}^{T} $, which leads to 
  \begin{align*}
 d_{\varphi }(\bsbSig^\circ)= \min_{\boldsymbol{v}\in\mathbb{R}^{m},\lVert\boldsymbol{v}\rVert_{2}=1} &\sum_{i}\varphi((\boldsymbol{y}_{i}^{T}\boldsymbol{v})^{2}-\boldsymbol{v}^{T}\boldsymbol{\Sigma}^{\circ}\boldsymbol{v})  \wedge\sum_{i}\varphi(-(\boldsymbol{y}_{i}^{T}\boldsymbol{v})^{2}+\boldsymbol{v}^{T}\boldsymbol{\Sigma}^{\circ}\boldsymbol{v}),\end{align*} and   $\varphi(t)=1_{\geq0}(t)$  corresponds to the notion of     matrix depth  in \cite{chen2018robust}. (The unit-rank reduction to a vector  $\bsbv$ is however incompatible with      imposing elementwise sparsity in covariance estimation; see Section \ref{subsec:slack}
of our companion paper  for a new idea of how to   define sparsity induced   depth and deepest $s$-sparse estimators.)

Similarly, we can introduce depth for    meta-regression with multiple outcomes. This could be  helpful to alleviate  the stringent normality assumption in meta-analysis. Assume there are $n$ studies with  $\bsbSig_i$   as the known within-study covariances: $\bsby_i = \bsbX_i \bsbb + \bsbeps_i + \bsb{\delta}_i$ ($1\le i \le n$), where  $\bsbeps_i \sim  \mathcal N(\bsb0, \bsbSig_i)$ are independent of $\bsb{\delta}_i\sim \mathcal N(\bsb0, \bsbSig)$.     Let $\bsbR_i = (\bsby_i - \bsbX_i\bsbb)(\bsby_i - \bsbX_i\bsbb)^T$. When  the between-study covariance $\bsbSig$   is of interest and $\bsbb$ is held fixed, we have      $\bsbT_i(\bsbSig) = (\bsbSig + \bsbSig_i)^{-1} (\bsbSigma + \bsbSigma_i - \bsbR_i) (\bsbSig + \bsbSig_i)^{-1}$,    which  again results in a symmetric $\mathcal G$.

\paragraph*{Projected triangle depth}
Consider projecting   all data points $\boldsymbol{z}_{i}\in\mathbb{R}^{m}$
($1\leq i\leq n$) to $\mathbb R^2$ to calculate the simplicial depth  \citep{liu1990}.
     Let $\triangle(\boldsymbol{z}_{i},\boldsymbol{z}_{j},\boldsymbol{z}_{k})$
denote the non-degenerate triangle formed by $\boldsymbol{z}_{i},\boldsymbol{z}_{j},\boldsymbol{z}_{k}$ and   assume that the data have been pre-processed to remove any collinearity.
Given any $\boldsymbol{V}\in\mathbb{R}^{m\times2}:\boldsymbol{V}^{T}\boldsymbol{V}=\boldsymbol{I}_{2\times2}$,
denote the projected point of $\boldsymbol{z}$   by $\Proj_{\boldsymbol{V}}(\boldsymbol{z})=\boldsymbol{V}^{T}\boldsymbol{z}\in\mathbb{R}^{2}$
and the augmented point by $\bar{\Proj}_{\boldsymbol{V}}(\boldsymbol{z})=[\boldsymbol{z}^{T}\boldsymbol{V}\;1]^{T}\in\mathbb{R}^{3}$.
Define the  \textit{projected} triangle depth for a target point $\boldsymbol{\mu}^{\circ}$
by $d(\boldsymbol{\mu}^{\circ})=\min_{\boldsymbol{V}}\#\{(i,j,k): i< j<k,  \Proj_{\boldsymbol{V}}(\boldsymbol{\mu}^{\circ})\in\triangle(\Proj_{\boldsymbol{V}}(\boldsymbol{z}_{i}), \Proj_{\boldsymbol{V}}(\boldsymbol{z}_{j}),\Proj_{\boldsymbol{V}}(\boldsymbol{z}_{k}))\}$  s.t. $ \boldsymbol{V}\in\mathbb{R}^{m\times2},\boldsymbol{V}^{T}\boldsymbol{V}=\boldsymbol{I}$,  and   introduce the $\varphi$-form\begin{equation}
\min_{\boldsymbol{V}\in\mathbb{R}^{m\times2}}\sum_{i<j<k}\prod_{l=1}^{3}\varphi(\xi_{l}^{\circ}(\boldsymbol{z}_{i},\boldsymbol{z}_{j},\boldsymbol{z}_{k};\boldsymbol{V}))\;\textrm{s.t.}\;\boldsymbol{V}^{T}\boldsymbol{V}=\boldsymbol{I},\label{eq:polished projected simplicial depth}
\end{equation}
where $\bsbxi ^{\circ}=[\xi_{l}^{\circ}]_{l=1}^{3}=[\bar{\Proj}_{\boldsymbol{V}}(\boldsymbol{z}_{i})\;\bar{\Proj}_{\boldsymbol{V}}(\boldsymbol{z}_{j})\;\bar{\Proj}_{\boldsymbol{V}}(\boldsymbol{z}_{k})]^{-1}\bar{\Proj}_{\boldsymbol{V}}(\boldsymbol{\mu}^{\circ})$.
Here, we used the fact that     $\Proj_{\boldsymbol{V}}(\boldsymbol{\mu}^{\circ})$
belongs to the projected triangle $\triangle(\Proj_{\boldsymbol{V}}(\boldsymbol{y}_{i}),\Proj_{\boldsymbol{V}}(\boldsymbol{y}_{j}),\Proj_{\boldsymbol{V}}(\boldsymbol{y}_{k}))$
if and only if $[\bar{\Proj}_{\boldsymbol{V}}(\boldsymbol{z}_{i})\ \ \allowbreak \bar{\Proj}_{\boldsymbol{V}}(z_{j})\   \allowbreak \ \bar{\Proj}_{\boldsymbol{V}}(\boldsymbol{z}_{k})]\boldsymbol{\xi}^{\circ}=\bar{\Proj}_{\boldsymbol{V}}(\boldsymbol{\mu}^{\circ})$
has a nonnegative solution $\boldsymbol{\xi}^{\circ}$. Because     $\bsbxi ^{\circ}$   is smooth in $\boldsymbol{V}$,     the  optimization techniques
developed in Section \ref{sec:Data-depth-computation} apply.  A similar
formulation can be given for the simplicial depth without
projection, and  to speed up the computation, one may   consider a randomized version as in
\cite{Afshani2009}.

\section{Optimization-based   Depth Computation}
\label{sec:Data-depth-computation}

\begin{figure}[htb]
\begin{centering}
\includegraphics[width=0.65\columnwidth,height=2.5in]{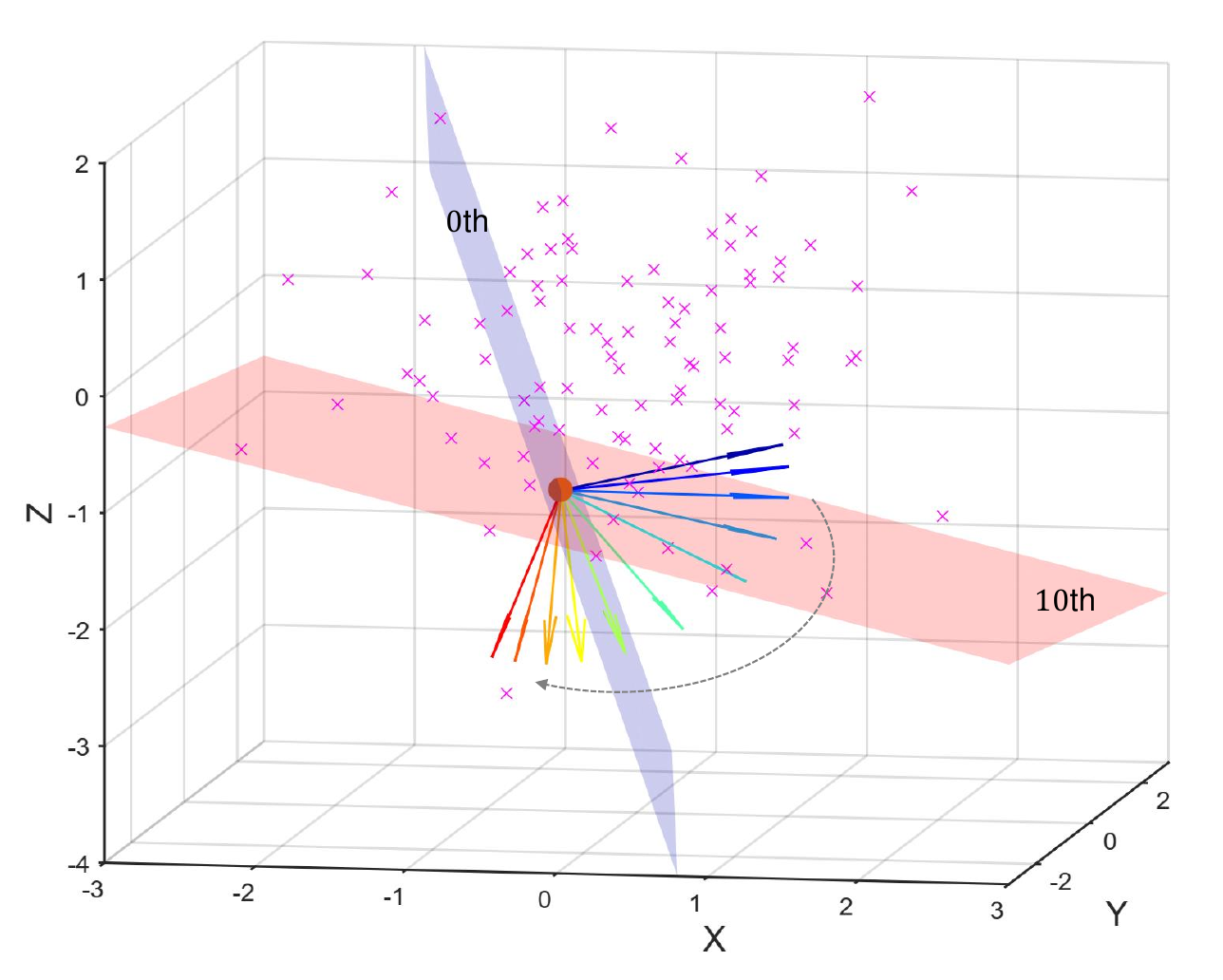}
\par\end{centering}
\caption{\small An example of optimization-based depth computation. The initial half-space  is in one of the worst directions, but after  $10$ steps, the optimal half-space
is found. \label{fig:Algorithm-progresses-PG APG 1}}
\end{figure}

 The    biggest obstacle to the application of    Tukey-type  depths   is perhaps the  heavy computational cost as mentioned in Section \ref{sec:intro}. Even in moderate dimensions,  the available methods  often suffer from either    prohibitively
high computational complexity or poor accuracy. Unlike many existing
algorithms and procedures that are designed based on geometry, or try to find smart  ways of numeration or  search,   this section    develops \textit{optimization} based depth computation with a rigorous convergence guarantee. Our ultimate target in this section is  $d_{01}$ but we will see that the $\varphi$-form  data depth facilitates algorithm design.   Before    describing   the  thorough detail,  it may
help the reader to check Figure \ref{fig:Algorithm-progresses-PG APG 1}
for an illustration of the power brought  by optimization. Even though
the initial   half-space  is in one of the worst directions,   the optimal half-space is found in $10$
iterations. An  outline of the associated algorithm is given in Appendix \ref{sec:algs}. 


For clarity, we will mainly use the polished half-space depth to describe  the derivation details, although in principle the same algorithm design      applies to the polished subspace depth as well. Because the loss in supervised learning is typically placed on the systematic component    $\bsbTh = \boldsymbol{X}\boldsymbol{B}$, and we denote by
 $\bar{l}(\boldsymbol{\Theta};\boldsymbol{Y})= \sum_{i}l_0(\boldsymbol{B}^T \boldsymbol{x}_{i}; \boldsymbol{y}_{i})/n$  the estimation   criterion with $\ell_0\in \mathcal C^{1}$. Then, the    depth  problem $\min \sum_{i}\varphi(\langle\boldsymbol{V},\boldsymbol{T}_{i}^{\circ}\rangle)\;\textrm{s.t.}\;\lVert\boldsymbol{V}\rVert_{F}=1, \boldsymbol{V}\in\mathcal G$ can  be restated in a trace form that is perhaps more amenable to matrix optimization: 
 \begin{equation}
\min_{\bsbV\in \mathcal G, \,  \lVert\boldsymbol{V}\rVert_{F}=1}f(\boldsymbol{V})\triangleq\textrm{Tr}\{\varphi(\boldsymbol{X}\boldsymbol{V}\boldsymbol{R}^{T})\},\label{eq:general regression data depth approximate opt}
\end{equation}
where $\varphi$ is applied  elementwise, i.e., $\varphi(\boldsymbol{X})_{ij}=\varphi(x_{ij})$,
and $$\boldsymbol{R}=\nabla_{\boldsymbol{\Theta}}\bar{l}\,\lvert_{\boldsymbol{\Theta}=\boldsymbol{X}\boldsymbol{B}^{\circ}}.$$
A   particular instance is the  GLM depth defined in (\ref{eq:polished regression data depth}), where   $\boldsymbol{R}=b^{\prime}(\boldsymbol{X}\boldsymbol{B}^{\circ})-\boldsymbol{Y}$. We can also write     $\boldsymbol{R} =[\boldsymbol{r}_1, \ldots, \boldsymbol{r}_n]^T$ with $\boldsymbol{r}_i=  \nabla l_0 (\vect(\boldsymbol{\Theta}[i,]);  \boldsymbol{y}_{i}) $, and  then $f(\bsbV) = \sum_i \varphi ( \langle \bsbV, \bsbT_i^{\circ}\rangle)$  and $\boldsymbol T_i^\circ =\boldsymbol{x}_{i}\boldsymbol{r}_{i}^{T} $. Formally, given $\bsbX^T \bsbR = \bsb0$, where the $i$th row  of $\bsbR$ depends on the $i$th sample  $(\bsbx_i , \bsby_i)$ only (thereby  sample-additive), the associated depth   objective is   $\textrm{Tr}\{\varphi(\boldsymbol{X}\boldsymbol{V}\boldsymbol{R}^{T})\}$.

Assume  that $\varphi$ is continuously differentiable    for now.
We can  develop a prototype algorithm following the principle of    majorization-minimization (MM)   \citep{Hunter2004tutorial}, where   a surrogate function needs to be created to majorize  the objective so that minimizing this surrogate function drives it downhill. We use   a quadratic surrogate function:
\begin{equation}
g_{\rho}(\boldsymbol{V},\boldsymbol{V}^{-})=f(\boldsymbol{V}^{-})+\langle\boldsymbol{X}^{T}(\textrm{diag}(\varphi^{\prime}(\boldsymbol{X}\boldsymbol{V}^-\boldsymbol{R}^{T}))\boldsymbol{R},\boldsymbol{V}-\boldsymbol{V}^{-}\rangle+\frac{\rho}{2}\lVert\boldsymbol{V}-\boldsymbol{V}^{-}\rVert_{F}^{2},\label{eq:surrogate}
\end{equation}
where $\rho>0$ and $\mbox{diag}(\bsbA) $ is a diagonal matrix formed by the diagonal entries of $\bsbA$. Here, $\boldsymbol{X}^{T}(\textrm{diag}(\varphi^{\prime}(\boldsymbol{X}\boldsymbol{V}^-\boldsymbol{R}^{T}))\boldsymbol{R}$ is   the gradient of $f$;  in implementation, the diagonal entries of $ \boldsymbol{X}\boldsymbol{V}\boldsymbol{R}^{T} $ can be  efficiently  calculated by   the row sums   of $(\bsbX \bsbV)\circ \bsbR$, where  $\circ$ denotes the elementwise product.
Starting with $\boldsymbol{V}^{(0)}:\lVert\boldsymbol{V}^{(0)}\rVert_{F}=1$,
we define a sequence of $\bsbV$-iterates by
\begin{equation}
\boldsymbol{V}^{(t+1)}\in\argmin_{\bsbV \in \mathcal G,\, \lVert\boldsymbol{V}\rVert_{F}=1}g_{\rho_{t}}(\boldsymbol{V},\boldsymbol{V}^{(t)}),\label{eq:iterates}
\end{equation}
for any $t\geq0$. We prove a convergence result for the resulting algorithm  assuming
$\varphi^{\prime}$ is Lipschitz continuous: $
\lvert\varphi^{\prime}(x)-\varphi^{\prime}(y)\rvert\leq L\lvert x-y\rvert$, $\forall x ,y\in\mathbb{R}$. Recall that $\| \cdot\|_2$ denotes the spectral norm of the enclosed matrix. 

\begin{thm}
\label{thm:gradient of f Lip continous}If $\rho_{t}$ is chosen large enough,  e.g.,
$\rho_{t}\geq L\lVert\boldsymbol{X}\rVert_{2}^{2}\lVert\boldsymbol{R}\rVert_{2}^{2}$,
then \eqref{eq:iterates} satisfies
\begin{equation}
f(\boldsymbol{V}^{(t+1)})\leq g_{\rho_{t}}(\boldsymbol{V}^{(t+1)},\boldsymbol{V}^{(t)})\leq g_{\rho_{t}}(\boldsymbol{V}^{(t)},\boldsymbol{V}^{(t)})=f(\boldsymbol{V}^{(t)}), \ \forall t\ge 0\label{eq:objective decreasing}
\end{equation}
That is, the objective function value is guaranteed non-increasing throughout the iteration.
\end{thm}

The convergence of the algorithm
holds more generally. The Lipschitz parameter is
 used to derive  a {universal} step-size; in implementation, we    recommend   performing  a line search. Specifically,  we can   decrease   $1/\rho_{t}$ until $f(\boldsymbol{V}^{(t+1)}(\rho_{t}))\leq g_{\rho_{t}}(\boldsymbol{V}^{(t+1)}(\rho_{t}),\boldsymbol{V}^{(t)})$
is satisfied (and so    $f(\boldsymbol{V}^{(t+1)})\leq f(\boldsymbol{V}^{(t)})$
  still holds for any $t$). The decrease in function value in the pursuit of   projection direction offers more stability  than geometry or search based algorithms. The surrogate  via linearization   applies to polished subspace depth as well. 

Because $g_{\rho}(\boldsymbol{V},\boldsymbol{V}^{-})=\rho \lVert\boldsymbol{V}-(\boldsymbol{V}^{-}- ({1}/{\rho})\boldsymbol{X}^{T}(\textrm{diag}(\varphi^{\prime}(\boldsymbol{X}\boldsymbol{V}^-\boldsymbol{R}^{T}))\boldsymbol{R})\rVert_{F}^{2}/2+ f(\boldsymbol{V}^{-}) - (1/2\rho) \|\boldsymbol{X}^{T}(\textrm{diag}(\varphi^{\prime}(\boldsymbol{X}\boldsymbol{V}^-\boldsymbol{R}^{T}))\boldsymbol{R}\|_F^2 $, the problem at each iteration boils down to
\begin{equation}
\min_{} \ \lVert\boldsymbol{V}-(\boldsymbol{V}^{(t)}-\frac{1}{\rho_{t}}\boldsymbol{G}^{(t)})\rVert_{F}^{2}\quad\textrm{s.t.\quad}\lVert\boldsymbol{V}\rVert_{F}=1, \boldsymbol{V}\in \mathcal G\label{eq:proximal like optimization}
\end{equation}
where  $\boldsymbol{G}^{(t)}=\boldsymbol{X}^{T}(\textrm{diag}(\varphi^{\prime}(\boldsymbol{X}\boldsymbol{V}^{(t)}\boldsymbol{R}^{T}))\boldsymbol{R}$.
Eqn. \eqref{eq:proximal like optimization}   has   many variants depending on the  projection space constraint. For instance, when  solving \eqref{psd-r} or \eqref{eq:polished projected simplicial depth}, the problem after linearization projects to  a Stiefel manifold   instead of a sphere. Some  more examples are given in Appendix \ref{sec:structuredprojs}.

We assume that  $\mathcal G$ is a linear subspace in the rest of the section (which includes the class of Riemannian depth  in our companion paper). Then \eqref{eq:proximal like optimization}  can be converted  to a case of  Procrustes
rotation.  Define a linear mapping   $\bsbB= \mathcal G(\bsbA)   $ such that  $\vect(\bsbB) = \mathcal P_{\mathcal G} \vect(\bsbA)$, where   $\mathcal P_{\mathcal G}$ is the orthogonal projection matrix onto subspace $\mathcal G$.   By writing $\vect(\boldsymbol{V}^{(t)}-  \boldsymbol{G}^{(t)}/{\rho_{t}} )= \mathcal P_{\mathcal G}\vect(\boldsymbol{V}^{(t)}-  \boldsymbol{G}^{(t)}/{\rho_{t}} ) + \mathcal P_{\mathcal G}^\perp  \vect(\boldsymbol{V}^{(t)}-  \boldsymbol{G}^{(t)}/{\rho_{t}} )$, we obtain
 $$\boldsymbol{V}^{(t+1)}=\mathcal G(\boldsymbol{V}^{(t)}-  \boldsymbol{G}^{(t)}/{\rho_{t}})/\lVert\mathcal G(\boldsymbol{V}^{(t)}-  \boldsymbol{G}^{(t)}/{\rho_{t}})\rVert_{F}.$$
Though not a proximity operator due to    {nonconvexity},
 the projection   guarantees global optimality in solving \eqref{eq:proximal like optimization}  (cf.  Lemma \ref{pertubres0}). 

Furthermore, we find that           Nesterov's   \textit{second  acceleration}  for convex programming  \citep{Nesterov2004book},  which
  attains the optimal convergence rate of $\mathcal O(1/t^2)$  among first-order methods, can   be modified  to   speed  the convergence of the prototype algorithm.
(Empirically, Nesterov's first acceleration   appears to be also effective, but we cannot provide its theoretical support.) To aid the presentation of the acceleration scheme,  we  define the \textit{generalized Bregman function} \citep{SheetalBreg} for
  any     continuously differentiable $\psi$
\begin{equation}
\bm{\Delta}_{\psi}(\bsbb,\bsbg) := \psi(\bsbb) - \psi(\bsbg) - \langle \nabla \psi(\bsbg), \bsbb-\bsbg \rangle.
\end{equation}
When $\psi$ is   strictly convex, $\bm{\Delta}_{\psi}$ becomes the standard Bregman divergence $\mathbf{D}_{\psi}(\bsbb,\bsbg)$ \citep{Bregman1967}. A simple example is $\mathbf{D}_2(\bsbb,\bsbg) :=  \|\bm\beta-\bm\gamma\|_2^2/2$, where   $\Breg_2$ denotes the Bregman associated with the half-squared-error-loss function, and its matrix version is $\Breg_2 (\bsbA, \bsbB) = \| \vect (\bsbA) - \vect(\bsbB)\|_2^2 /2 = \| \bsbA - \bsbB\|_F^2/2$. In general, $\bm{\Delta}_\psi$ or $\mathbf{D}_\psi$ may not be symmetric.

Consider the following {momentum}-based update    which involves three major sequences $\bsbU^{(t)}$, $\bsbW^{(t)}$, $\bsbV^{(t)}$, $t=0, 1, \ldots$ (starting with $\theta_0=1$ and any  $\bsbW^{(0)}\in \mathbb R^{p\times m}$):
\begin{align}
\bsbU^{(t)} &= (1-\theta_t)\bsbV^{(t)} + \theta_t\bsbW^{(t)}, \label{acc2-alg1}\\
\boldsymbol{G}^{(t)}&=\boldsymbol{X}^{T}(\textrm{diag}(\varphi^{\prime}(\boldsymbol{X}\boldsymbol{U}^{(t)}\boldsymbol{R}^{T}))\boldsymbol{R},\\
\bsbXi^{(t)} & =\mathcal G( \bsbW^{(t)}  - \boldsymbol{G}^{(t)}/\{\theta_t\rho_t \}), \label{Xiupd}\\
\bsbW^{(t+1)} &= \bsbXi^{(t)} / \|\bsbXi^{(t)}\|_F,
\label{acc2-alg2}\\
\bsbV^{(t+1)} &= (1-\theta_t)\bsbV^{(t)} + \theta_t\bsbW^{(t+1)}. \label{acc2-alg3}
\end{align}

The design of the relaxation parameters $\theta_t$ and inverse stepsize parameters $\rho_t$ holds the key to acceleration.
We propose the following line search criterion
\begin{align}
&R_t \triangleq \theta_t^2\rho_t\mathbf{D}_2 (\bsbW^{(t+1)},\bsbW^{(t)}) - \bm\Delta_{f}(\bsbV^{(t+1)},\bsbU^{(t)}) + (1-\theta_t)\bm\Delta_{f}(\bsbV^{(t)},\bsbU^{(t)})
 \ge 0,~~~~~\label{acc2_search1}\\
&\frac{\theta_t^2}{1-\theta_t} = \frac{\rho_{t-1}\theta_{t-1}^2}{\rho_t},~\theta_t \ge 0, \rho_t > 0, t\ge 1.\label{acc2_search2}
\end{align}
and $ \theta_0 = 1$.
Some implementation details     are given in Algorithm \ref{alg:ap}. When $f$  has $L$-Lipschitz continuity in its gradient, \eqref{acc2_search1} is implied by
\begin{equation}
\theta_t^2(\rho_t - L )\mathbf{D}_2 (\bsbW^{(t+1)},\bsbW^{(t)}) + (1-\theta_t)\bm\Delta_{f}(\bsbV^{(t)},\bsbU^{(t)}) \ge 0.
\end{equation}
If, further,   $f$ is convex, taking   $\rho_t = L$ and $\theta_{t+1} = (\sqrt{\theta_t^4 + 4\theta_t^2} - \theta_t^2)/2$ gives the standard   convex second acceleration  \citep{Tseng2010}.
The reasonability of \eqref{acc2_search1} in our nonconvex setup can be seen from the following theorem, where the  convergence is shown under  a proper  discrepancy measure.

\begin{thm}\label{th:comp_acc2}
Given any $\rho_t>0$ ($t\ge 0$), consider the algorithm defined by \eqref{acc2-alg1}--\eqref{acc2-alg3} and \eqref{acc2_search2}. Then for \emph{any} $\bm{V} \in \mathcal G:  \| \bsbV\|_F=1$ and $T\ge 0$,
\begin{equation} \label{acc2_result}
\begin{split}
\frac{f(\bm{V}^{(T+1)})-f(\bm{V})}{\theta_T^2\rho_T} + T\cdot\mathop{\avg}_{0\le t\le T}\frac{\mathcal E_t(\bm V)}{\theta_t\rho_t} + T\cdot\mathop{\avg}_{0\le t\le T}\frac{R_t}{\theta_t^2\rho_t} \\ \le \mathbf{D}_2 (\bm{V},\bm{W}^{(0)}) - \mathbf{D}_2 (\bm{V},\bm{ W}^{(T+1)}),\end{split}
\end{equation}
where $\mathcal E_t(\bm V) = \bm\Delta_{f}(\bm{V},\bm{U}^{(t)}) +   \theta_t\rho_t( \| \bsbXi^{(t)}\|_F-1)\mathbf{D}_2   (\bsbV, \bsbW^{(t+1)})$.
\end{thm}

Typically,  \eqref{acc2_search1}  
involves a line search. If the condition fails for the current value of $\rho_t$,  one can set $\rho_t = \beta\rho_t$ for some $\beta > 1$ (say $2$) and recalculate $\theta_t$  according to \eqref{acc2_search2} and other quantities defined in \eqref{acc2-alg1}--\eqref{acc2-alg3} to verify  \eqref{acc2_search1}  
 again. Moreover, if $\rho_t/\rho_{t-1}\ge 1-(at+ab+1)/(t+b-1)^2$ for some constants $a,b$: $a\ge 0, b\ge a+1$, say, $\rho_t/\rho_{t-1}\ge 1-1/t^2$, then by induction, it is easy to show $\theta_t \le (a+2)/(t+b) = \mathcal O(1/t)$, and so
\[
\theta_T^2 = \mathcal O(1/T^2) \mbox{~~and~~} \sum_{0\le t\le T} 1/(\rho_t\theta_t)  \ge \mathcal O(T^2/\rho_{T}).
\]
Hence with $\sum_{t=0}^T R_t/(\theta_t^2\rho_t) \ge 0$ which is guaranteed by  $R_t\ge 0$, \eqref{acc2_result} implies
$f(\bsbV^{(T+1)})-f(\bsbV^\star) + \min_{0\le t\le T}\mathcal E_t(\bsbV^\star) \le \mathcal O( \rho_T/T^2)$ for any optimal solution $\bsbV^\star$.  If  $R_t\ge 0$  does not hold  after a prescribed number $M$ of searches, we can  pick the $(\rho_t, \theta_t)$  giving   the largest $R_t/(\theta_t^2 \rho_t)$ in view of Theorem \ref{th:comp_acc2}. Experience shows that  the momentum-based  update always  speeds the convergence.

To initialize the algorithm, we adopt a simple but effective multi-start strategy by
\cite{Rousseeuw1998}:
select $n_0$   observations at random, and  for each observation  calculate   $-\boldsymbol{x}_{i}\boldsymbol{r}_{i}^{T}$
 as  a candidate direction. We  suggest adding the direction from spherical PCA \citep{locantore1999robust}.
     Section \ref{sec:Simulation-Studies} uses       $n_0=10$.
  Compared with other methods,  our algorithm   is much less demanding on the initial value (cf.    Figure \ref{fig:Algorithm-progresses-PG APG 1}
and Table \ref{simulation:regression}).

The efficient algorithm for    polished depth can be used to obtain  $d_{01}$. A simple  means is by {successive} optimization as in  interior point methods \citep{Boyd2004}. Concretely,  use a series of functions to approximate
$1_{\ge 0}(t) $ or $\sgn(t)$ (such as  $\varphi_\zeta (x) =   \Phi(\zeta x)$ with $\Phi$ the standard normal distribution,
$
\tanh(\zeta x)= ({e^{\zeta x}-e^{-\zeta x}})/({e^{\zeta x}+e^{-\zeta x}})
$,
or $(2/\pi)\arctan(\zeta x)$)  
and solve $\min_{ \lVert\boldsymbol{V}\rVert_{F}=1}\textrm{Tr}\{\varphi_{\zeta}(\boldsymbol{X}\boldsymbol{V}\boldsymbol{R}^{T})\}$
  with $\zeta\rightarrow\infty$.
   Fortunately, the finite number of data points often means    a finitely large  $\zeta$  suffices in implementation.    The resultant algorithm, referred to as the successive accelerated projection (\textbf{{SAP}}),  is  summarized   in    Appendix \ref{sec:algs}. It has implementation ease,  and shows remarkable  improvement     over     existing algorithms  in  accuracy and computational time (especially when $m\ge 20 $). 

\begin{remark}[Nested algorithm design for computing composite depth] \label{rem:depth median computation}

Suppose that an  event of interest is given by    $  \Omega_0$ as a subset of $\Omega$, and the goal is to assess its reliability. The previous algorithms studying a simple hypothesis  (assuming  $\Omega_0$ is a singleton)  can be possibly adapted to the general case. 

Concretely, for testing $H_0:\bsbB\in \Omega_0$,  we define the \emph{``composite depth''} of $\Omega_0$ by
   \begin{align}
d_{01}(\Omega_0) =    \max_{\boldsymbol{B}\in\Omega_0}d_{01}(\boldsymbol{B}) . \label{eqcompositedep}
   \end{align}
    In the extreme case    $\Omega_0 =\mathbb R^{p\times m}$, \eqref{eqcompositedep} amounts to finding the deepest estimate.
  How to estimate the deepest point is a challenging topic  beyond the scope of
the current paper, but motivated by Danskin's theorem  \citep{Bert}, the
algorithms in this section can be incorporated into  a \emph{nested} algorithm
for solving the nonconvex ``max-min''  optimization problem   $\max_{\boldsymbol{B}\in\mathbb{R}^{p\times m}}d_{\varphi}(\boldsymbol{B})$
or \[
\max_{\boldsymbol{B}\in\mathbb{R}^{p\times m}}\min_{\lVert\boldsymbol{V}\rVert_{F}=1}f(\boldsymbol{B},\boldsymbol{V})\triangleq\textrm{Tr}[\varphi(\boldsymbol{X}\boldsymbol{V}
\{\boldsymbol{R}(\boldsymbol{X}\boldsymbol{B})\}^{T})].
\]
Specifically, assuming that $\varphi$
is smooth (otherwise employ  a successive optimization scheme as before) and
 $\bsbR(\bsbTh) =[R_{ik}(\theta_{ik})]$,  apply    the chain rule:
$
\nabla_{\boldsymbol{B}}f(\boldsymbol{B},\boldsymbol{V})=\boldsymbol{X}^{T}\{\nabla_{\boldsymbol{\Theta}}\boldsymbol{R}(\boldsymbol{\Theta}) \circ \textrm{[diag}(\varphi^{\prime}(\boldsymbol{X}\boldsymbol{V}\boldsymbol{R}^{T}))\boldsymbol{X}\boldsymbol{V}] \}$ where    $\nabla_{\boldsymbol{\Theta}}\boldsymbol{R}(\boldsymbol{\Theta})=[  R_{ik}' (  \theta_{ik})]\in\mathbb{R}^{n\times m}$.
Then, given $\bsbV(\boldsymbol{B}^{(t)})$  as a solution to $\min_{\boldsymbol{V}: \|\bsbV\|_F=1}  \allowbreak f(\boldsymbol{B}^{(t)},\boldsymbol{V})$, the $\bsbB$-update    is    $\boldsymbol{B}^{(t+1)}=\boldsymbol{B}^{(t)}+\alpha_{t}\nabla_{\boldsymbol{B}}f(\boldsymbol{B}^{(t)},\boldsymbol{V}(\bsbB^{(t)}))$, where $\alpha_{t}$ is the step-size that can be determined by say Armijo  line search.
Although it is difficult to provide any provable guarantee due to  the lack of convexity for our max-min problem, the above algorithm appears to work in practice. For     $\max_{\boldsymbol{B}\in\Omega_0 }\allowbreak\min_{\lVert\boldsymbol{V}\rVert_{F}=1}f(\boldsymbol{B},\boldsymbol{V})$, one just needs to replace the gradient descent by projected gradient descent.
 In this way,   we can use  composite depth  to evaluate the data centrality of an event. The   influence-driven nonasymptotic  index can serve  as a surrogate for the $p$-value, without  making  any distributional or large-sample assumptions.
\end{remark}
\section{Experiments}
\label{sec:Simulation-Studies}

This part generates synthetic data to compare    some popular   methods and  SAP in  location  and regression depth computation. To meet   the challenges of modern data applications,  our setups have higher dimensions than  most existing works (where  a dimension lower than 10 is often used). The evaluation metrics are, naturally,
the value of depth (the objective function value of the  associated minimization problem with $\varphi = 1_{\ge 0}$) and   computation time
(in CPU seconds), both averaged over $50 $ runs. An excellent   algorithm should show reasonably   low depth   and  computational complexity. Since    scalability is a major concern, we will vary the   problem dimensions in most experiments.  In running SAP, the termination criterion is met if      the change in objective     is less than 1e-2, the max-norm of  the gradient is less than 1 or the number of iterations exceeds 5000.
     As aforementioned, in all the  SAP experiments, we just used 10 random starting points.  All simulation experiments were
performed with Matlab 2018a on a machine with  Intel Core I5-4460S and 16GB RAM.

\paragraph{Location depth}

In the first setting, the observations are generated by $z_{ij}\stackrel{\text{i.i.d.}}{\sim}\mathcal{N}(0,1)$
with $n=100$,  $m=10, 20, 30, 40$, and the target point is   $\boldsymbol{\mu}^{\circ}=[0.1,  \ldots, 0.1]^{T}$.
Due to the \textit{curse of dimensionality}, $\bsbmu^{\circ}$ should behave more and more like a boundary point as  $m$ increases.  Table \ref{simulation:location:n100} shows  a performance comparison between   SAP and     some methods implemented in      {R} packages  \texttt{ddalpha} \citep{ddalpha2016},  \texttt{depth} \citep{depth2017},
and \texttt{DepthProc} \citep{DepthProc2017} and  \texttt{MTMSA} \citep{shao2020computing}. In calling the first three packages, we used the ``approximate'' option (since no algorithm can compute the exact depth when $m>6$) and increased the number of initial random directions from the default 1000 to 20,000 to boost their accuracy; the other parameters are taken their default values. The implementation of the continuous  \texttt{MTMSA} has four recommended configurations. We reported the results of scheme II in their paper since it consistently gave lower depth values than the other three in our experiments.

When $m=10$, all methods gave similar depth values.  But when $m=40$, SAP showed a significantly lower depth  than the other methods.  Our algorithm  is also  the winner in terms of computational scalability.
\begin{table}[h]
{\footnotesize{
\caption{ Location depth   comparison between \texttt{ddalpha},   \texttt{depth},
 \texttt{DepthProc}, and SAP in setting 1 ($n=100$).   \label{simulation:location:n100}}
\begin{center}
\begin{tabular}{l cc cc cc cc} \hline
& \multicolumn{2}{c}{$m=10$} & \multicolumn{2}{c}{$m=20$} & \multicolumn{2}{c}{$m=30$} & \multicolumn{2}{c}{$m=40$} \\
\hline & Time & Depth  & Time & Depth  & Time & Depth  & Time & Depth \\
\texttt{ddalpha}    & $0.04$ & $0.28$ & $0.05$  & $0.27$& $0.07$ & $0.25$ & $0.11$ & $0.25$ \\
\texttt{depth}     & $0.27$ & $0.27$ & $1.1$  & $0.22$& $2.7$ & $0.18$ & $5.6$ & $0.15$ \\
 \texttt{DepthProc}     & $3.3$ & $0.27$ & $3.4$  & $0.27$& $3.4$ & $0.24$ & $3.4$ & $0.24$ \\
\texttt{MTMSA}    & $0.25$ & $0.24$ & $0.31$ & $0.18$ & $0.37$ &$0.14$ &$0.43$ &$0.13$ \\
\textbf{SAP}   &$0.02$ & $0.22$ & $0.02$ & $0.14$& $0.02$ & $0.09$ & $0.02$ & $0.06$\\
\hline
\end{tabular}
\end{center}
}}
\end{table}

\begin{figure}[htb!]
\begin{centering}
\includegraphics[width=.7\columnwidth, height=2.8in]{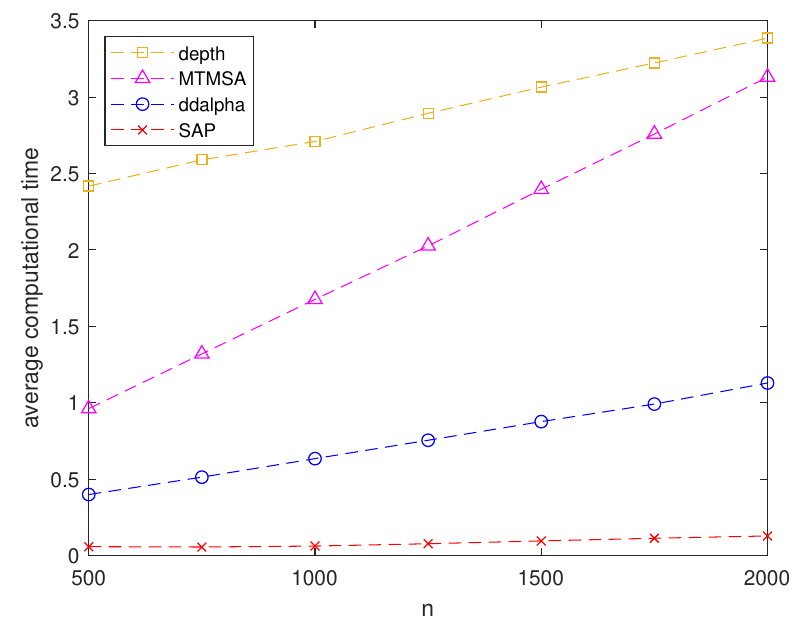}
\caption{\small Computational time comparison  between \texttt{ddalpha}, \texttt{depth},  \texttt{MTMSA} and SAP, averaged over 50 runs,  as a function of $n$. (\texttt{DepthProc} is not included due to its  high cost.)   \label{fig:comptime} }
\end{centering}
\end{figure}

In setting 2,  the number of observations is increased to $n=1\mbox{,}000$,  the other parameters remaining the same. We also performed a scalability experiment with increasing  values of $n$,  in terms of   computational time.  In setting 3,  $z_{ij}\stackrel{\text{i.i.d.}}{\sim}U(-3,3)$,
 $n=500$,  $m=50$, and the target point  $\boldsymbol{\mu}^{\circ}$   varies.  In these experiments, the package   \texttt{DepthProc} was unstable and  prone to crashing.    The results are summarized in    Table \ref{simulation:location:n1000}, Figure \ref{fig:comptime}, and Table \ref{simulation:location:setting3}, and similar conclusions can be drawn. It is worth mentioning that getting very similar depth values is not necessarily a sign of accuracy. In fact, because these different methods solve the same  $\bsbV$-minimization problem with      depth  as the objective function value, we favor the half-space direction $\hat \bsbV$ that  gives the lowest depth. Overall, our   optimization-assisted half-space depth computation  brings         substantial improvements in accuracy, complexity and initialization.

\begin{table}[h]
{\footnotesize{
\caption{ Location depth   comparison between \texttt{ddalpha},   \texttt{depth},
 \texttt{DepthProc}, and SAP in setting 2 ($n=1000$).   \label{simulation:location:n1000}}
\begin{center}
\begin{tabular}{lcccccccc}
\hline
           & \multicolumn{2}{c}{$m=10$} & \multicolumn{2}{c}{$m=20$} & \multicolumn{2}{c}{$m=30$} & \multicolumn{2}{c}{$m=40$} \\ \hline
           & Time       & Depth       & Time       & Depth       & Time       & Depth       & Time       & Depth       \\
\texttt{ddalpha}    & 0.41       & 0.37        & 0.55       & 0.35        & 0.69       & 0.34        & 0.98       & 0.34        \\
\texttt{depth}      & 0.50       & 0.37        & 1.4        & 0.35        & 3.1        & 0.34        & 6.3        & 0.34        \\
\texttt{DepthProc}  & 6.4        & 0.37        & 6.5        & 0.35        & 6.6        & 0.34        & 6.7        & 0.34        \\
 \texttt{MTMSA}      & 1.0       & 0.35        & 1.3       & 0.31        & 1.6       & 0.28        & 2.0       & 0.26        \\
 \textbf{SAP} & $0.03$       & $0.34$        & $0.05$       & $0.28$        & $0.06$       & $0.23$        & $0.08$       & $0.20$         \\
 \hline
\end{tabular}
\end{center}
}}
\end{table}

\begin{table}[h]
{\footnotesize{
\caption{ Location depth   comparison between \texttt{ddalpha},   \texttt{depth},
 \texttt{DepthProc}, and SAP in setting 3 with different locations of interest ($n=500, m= 50$). \label{simulation:location:setting3}}
\begin{center}
\begin{tabular}{lcccccc}
\hline
          & \multicolumn{2}{c}{$ \mu_j^\circ= 0$} & \multicolumn{2}{c}{$\mu_j^\circ\!\sim\!\mathcal N( 0,  0.1^2\!)$} & \multicolumn{2}{c}{ $\mu_j^\circ\!\sim\!U (\!-.5, .5\!)\!$}  \\ \hline
           & Time         & Depth        & Time             & Depth            & Time          & Depth         \\
\texttt{ddalpha}    & 0.47         & 0.41         & 0.39          & 0.35                & 0.39             & 0.23     \\
\texttt{depth}      & 8.2          & 0.38         & 8.2           & 0.34                & 8.1              & 0.23                      \\
\texttt{DepthProc}  & 7.7          & 0.41         & 5.4              & 0.35             & 5.4              & 0.23                     \\
\texttt{MTMSA}      & 1.2         & 0.37         & 1.3           & 0.28                & 1.5              & 0.10     \\
\textbf{SAP} & $0.15$         & $0.25$                & $0.10$          & $0.17$                & $0.06$             & $0.04$          \\  \hline
\end{tabular}
\end{center}
}}
\end{table}

\paragraph{Regression depth}

Here, we compute regression depth with   SAP and a popular
package \texttt{mrfDepth} \citep{mrfDepth2017},   denoted
by MD below. The data are generated according to  $y_{i}=\sum_{j}x_{ij}\beta_{j}^{*}+\beta_{0}^{*}+\epsilon_{i}$
where   $x_{ij}\stackrel{\text{i.i.d.}}{\sim}\mathcal{N}(0,1)$,
 $\epsilon_{i}\stackrel{\text{i.i.d.}}{\sim}\mathcal{N}(0,1)$, $\bsbb^* = [\beta_0^*, \beta_{1}^{*}, \ldots, \beta_p^*]^T = [1, 1, \ldots, 1]^T$,  $n= 1000$ and  $p=10, 20, 30, 40$. We set    $\boldsymbol{\beta}^{\circ}=[0, 0, \ldots, 0]^{T}$ and anticipate it to  be  further away from the center of the data as  $p$ grows.

\begin{table}[h]
{\footnotesize{ \caption{ Regression depth: comparison between \texttt{mrfDepth}  (MD) and  SAP with Gaussian noise. Here, $n_0$ is the number of starting  points for each algorithm.\label{simulation:regression}}
\begin{center}
\begin{tabular}{l c cc cc cc cc} \hline
&$n_0$ &  \multicolumn{2}{c}{$p=10$} & \multicolumn{2}{c}{$p=20$} & \multicolumn{2}{c}{$p=30$} & \multicolumn{2}{c}{$p=40$} \\
\hline &  & Time & Depth  & Time & Depth  & Time & Depth  & Time & Depth \\
MD  & $250p$      & $0.24$ & $0.16$ & $0.7$  & $0.22$& $1.73$ & $0.27$ & $4.14$ & $0.29$ \\
MD & $50000p$             & $40.4$ & $0.11$ & $127.7$  & $0.17$& $329.4$ & $0.21$ & $774.3$ & $0.24$ \\
\textbf{SAP} & $10$   & $0.06$ & $0.09$ & $0.06$ & $0.06$& $0.07$ & $0.04$ & $0.07$ & $0.03$ \\
\hline
\end{tabular}
\end{center}
}}
\end{table}

By default, MD uses $n_0=250 p$  starting points by random sampling.   But it showed poor performance  in Table \ref{simulation:regression} (say   when   $p=40$).   In order to see the true potential of MD, we   enlarged  $n_0$ to   $50000    p$.   The extensive sampling  took much longer    time but  led to only a minor improvement. In fact,   the depth   computed by MD is monotonically increasing  in $p$ (from $0.16$ to $0.29$ when $n_0=250 p$, and   $0.11$  to $0.24$ when $n_0=50000 p$),  suggesting  the inherent difficulty of searching  in higher dimensions.

In comparison, our SAP algorithm  showed a correct decreasing trend, and  gave consistently lower depths  by use of only $10$ random starts. What is particularly impressive  is its computational cost---all SAP computations were done within $1$ second.

A similar experiment with Cauchy noise   $\epsilon_{i}\stackrel{\text{i.i.d.}}{\sim} {C}(0,1)$ was carried out in Table \ref{simulation:regression:cauchynoise} and our findings are the same.

\begin{table}[h]
{\footnotesize{ \caption{ Regression depth  comparison between \texttt{mrfDepth}    and  SAP under Cauchy noise.  \label{simulation:regression:cauchynoise}}
\begin{center}
\begin{tabular}{l c cc cc cc cc} \hline
&  &  \multicolumn{2}{c}{$p=10$} & \multicolumn{2}{c}{$p=20$} & \multicolumn{2}{c}{$p=30$} & \multicolumn{2}{c}{$p=40$} \\
\hline & $n_0$ & Time & Depth  & Time & Depth  & Time & Depth  & Time & Depth \\
MD  &     $250p$   & $0.24 $ & $0.22 $ & $0.74 $  & $ 0.27$& $1.89 $ & $0.29 $ & $4.41 $ & $0.31 $ \\
MD &     $50000p$         & $35.0$ & $0.19$ & $106.7$  & $0.22$& $271.6$ & $0.25$ & $616.2$ & $0.27$ \\
\textbf{SAP} & $10$   & $0.22$ & $0.17$ & $0.46$ & $0.13$& $0.50$ & $0.12$ & $0.69$ & $0.10$ \\
\hline
\end{tabular}
\end{center}
}}
\end{table}

\section{Summary}
Tukey's half-space depth considers all half-spaces that contain $\bsbmu^\circ$ in their boundaries \textit{or} in their interiors. A candidate half-space with normal direction $\bsbv$   can be characterized by  $\langle \bsbv, \bsbmu^\circ\rangle\ge 0$, and Tukey minimizes the number of observations belonging to the ``positive class''  $\langle \bsbv, \bsbz_i\rangle\ge 0$ to get an optimal half-space. In the location setup, the minimization implies that one only  needs to focus on  $\bsbv:  \langle \bsbv, \bsbmu^\circ\rangle= 0$,  the half-spaces containing $\bsbmu^\circ$ in the boundaries, so $\langle \bsbv, \bsbz_i\rangle\ge 0$ becomes  $\langle \bsbv, \bsbz_i-\bsbmu^\circ\rangle\ge 0$, and the objective  equivalent to the ``contrast'' $\#\{\bsbz_i: \langle \bsbv, \bsbz_i-\bsbmu^\circ\rangle\ge 0\} - \#\{\bsbz_i: \langle \bsbv, \bsbz_i-\bsbmu^\circ\rangle< 0\}$ as a relaxed, robust measure of how the underlying normal  equation of $\sum (\bsbz_i-\bsbmu) = \bsb0$ is obeyed. Polished subspace depth generalizes  $\bsbz_i-\bsbmu^\circ$  to an influence, confines  $\bsbv$  in the associated influence space, explores some possibilities of  ``soft''  classification and redescending measures,   generalizes the straight-line projection to an $r$-dimensional subspace projection, and discusses how to maintain  invariance in the new general setup.
The resulting \textbf{Tukeyfication} process applies broadly. The boundary restriction is often   without any loss of generality  (especially when $\mathcal G$ is the full Euclidean space); yet   there are cases where one wants to include the  interiors. See   Remark \ref{rmk:spdepth}  in  \cite{SheDepthII2021}, as well as  an ``order-2'' Tukeyfication when the loss is nonconvex or the constraint region is compact.

Our new matrix formulation of the problem facilitates   optimization algorithm design.  We utilized  linearization,     iterative Procrustes rotations, and Nesterov's momentum-based acceleration to develop  efficient algorithms with a convergence guarantee.  The experiments demonstrated the impressive performance  of optimization-based depth computation in accuracy, complexity and initialization.

Data depth can be used for nonparametric inference by exploiting data centrality
with no rigid model or presumed distribution assumptions.
   Tukeyfication  can also upgrade an ordinary method of estimation to a distribution-free, robust deepest estimation that can tolerate gross outliers. On the other hand, modern applications in high dimensional statistics and machine learning  often involve problems that are defined in a restricted space or  have nondifferentiability issues, for which the notion of depth needs to be carefully  calibrated and examined  \citep{SheDepthII2021}.


%
%


\appendix
\section{Algorithm summary }
\label{sec:algs}
The following algorithm is for computing polished half-space depth.

\begin{algorithm}[H]
\caption{\small Accelerated projection   for computing   \eqref{eq:general regression data depth approximate opt} with a   $\varphi\in \mathcal C^1$ \label{alg:ap}}
\textbf{Input} { $\varphi, \mathcal G, \boldsymbol{X},  \boldsymbol{R} $ (cf. \eqref{eq:general regression data depth approximate opt}) and     $\boldsymbol{W}^{(0)}$,  an  initial direction.   (Other parameters for line search:   $\rho_{\textrm{min}}>0$, $\beta>1$, $M\in \mathbb N$, e.g.,  $\rho_{\textrm{min}} = 1$, $\beta = 2$, $M=3$)
}
\begin{algorithmic}[1]
\State  $\theta_0 \leftarrow 1$, $t \leftarrow 0$; 
\While{not converged}
\State $\rho_{t}\leftarrow\rho_{\min}/\beta$, $s\leftarrow 0$
\Repeat
\State $s \leftarrow s+1$
\State $\rho_t \leftarrow \beta \rho_t$
\State  {if} $t\ge  1$, then $\theta_t=(\theta_{t-1}\sqrt{\rho^2_{t-1}\theta^2_{t-1}+4\rho_t\rho_{t-1}}-\rho_{t-1}\theta^2_{t-1})/2\rho_t $
\State $\bsbU^{(t)} \leftarrow (1-\theta_t)\bsbV^{(t)} + \theta_t\bsbW^{(t)}$
\State $\boldsymbol{G}^{(t)}\leftarrow\boldsymbol{X}^{T}(\textrm{diag}(\varphi^{\prime}(\boldsymbol{X}\boldsymbol{U}^{(t)}\boldsymbol{R}^{T}))\boldsymbol{R}$
\State $\bsbXi^{(t)} \leftarrow   \mathcal G ( \bsbW^{(t)}  -   \boldsymbol{G}^{(t)}/\{\theta_t\rho_t \})$
\State $\bsbW^{(t+1)} \leftarrow \bsbXi^{(t)} / \|\bsbXi^{(t)}\|_F$
\State $\bsbV^{(t+1)} \leftarrow (1-\theta_t)\bsbV^{(t)} + \theta_t\bsbW^{(t+1)}$
\State $R_t \leftarrow \theta_t^2\rho_t\mathbf{D}_2 (\bsbW^{(t+1)},\bsbW^{(t)}) - \bm\Delta_{f}(\bsbV^{(t+1)},\bsbU^{(t)}) + (1-\theta_t)\bm\Delta_{f}(\bsbV^{(t)},\bsbU^{(t)})$

\Until{$R_t\ge 0$ or $s>M$}
\State $t \leftarrow t+1$
\EndWhile
\State \Return $\boldsymbol{V}^{(t)}$.

 \end{algorithmic}
\end{algorithm}

The   algorithm  of {successive accelerated projection} (\textbf{SAP}) for computing $d_{01}$ (cf. \eqref{eq:general regression data depth approximate opt} with $\varphi =1_{\ge 0} $) runs as follows: start with     $\zeta\leftarrow1$, $\bsbV \leftarrow \bsbV^{(0)}$; 
repeat $\bsbV \leftarrow$  Algorithm \ref{alg:ap}  with  $\varphi_{\zeta}$, $\mathcal G$, $\boldsymbol{X}$, $\boldsymbol{R}$, $\bsbV$ as the input,
and update   $\zeta \leftarrow \alpha\zeta$,
until   $\zeta\leq\zeta_{\max}$. Here, $\boldsymbol{V}^{(0)}$ is an initial direction and  $\zeta_{\max}, \alpha$ are annealing parameters,  e.g., $\zeta_{\max} = 10$, $\alpha=1.25$.


\section{Structured projections}
\label{sec:structuredprojs}
Given a general matrix $\bsbA$, with $\bsbU_A \bsbD_A \bsbV_A^T$ as its reduced SVD, define $\bsbA^+ = \bsbV_A \bsbD_A^{-1} \bsbU_A^T$, ${\mathcal P}_{\bsbA} = \bsbU_A \bsbU_A^T$ and  ${\mathcal P}_{\bsbA}^{\perp}  = \bsbI -  P_{\bsbA}$.
 Define $0/0:=0$.

 \begin{lemma} \label{lem:spar0} For
$
\min_{\boldsymbol{V}\in\mathbb{R}^{p\times r}}\lVert\boldsymbol{Y}-\boldsymbol{V}\rVert_{2}^{2}\;\textrm{s.t.}\; \bsbV^T \bsbV  =\bsbI_{r\times r},\;
\bsbV \in \mathcal G$
where $\mathcal G$ is a subspace given by $ \{\bsbA \bsbC \bsbB^T: \forall \bsbC \}$, a globally optimal solution is $\mathcal G(\bsbY) \{\mathcal G(\bsbY)^T \mathcal G(\bsbY)\}^{-1/2}$, where $\mathcal G(\bsbY)  = \mathcal P_{\bsbA} \bsbY \mathcal P_{\bsbB}$.
\end{lemma}

The proof is omitted. This   subspace constrained Procrustes rotation is often useful in computing polished subspace depth.

 \begin{lemma}\label{lem:zeroout}
For  $\min_{\bsbv\in \mathbb R^p} \| \bsby - \bsbv\|_2^2$ s.t. $\|\bsbv\|_2^2=1, \bsbA\bsbv_{\Omega} = \bsba$ with $\bsba\in {\mathcal P}_{\bsbA}$ and $\|\bsbA^+\bsba \|_2\le 1$, the  globally optimal solution is $\bsbA^+ \bsba + {\mathcal P}_{\bsbA^T}^\perp \bsby (1 - \|\bsbA^+\bsba \|_2^2)^{1/2}/\|{\mathcal P}_{\bsbA^T}^\perp \bsby\|_2$.

 In particular,    for $\min_{\bsbv\in \mathbb R^p} \| \bsby - \bsbv\|_2^2$ s.t. $\|\bsbv\|_2^2=1, \bsbv_{\Omega} = \bsb0$, where $\Omega\subset\{1, \ldots, p\}$, the   optimal solution $\bsbv^\star$ satisfies $\bsbv^\star_\Omega=\bsb0$ and $\bsbv^\star_{\Omega^c} = \bsby_{\Omega^c} / \| \bsby_{\Omega^c}\|_2$. \end{lemma}
The proof is omitted.

\begin{lemma} \label{lem:spar1} For
$
\min_{\boldsymbol{v}\in\mathbb{R}^{p}}\lVert\boldsymbol{y}-\boldsymbol{v}\rVert_{2}^{2}\;\textrm{s.t.}\;\lVert\boldsymbol{v}\rVert_{2}=1,\;
\lVert\boldsymbol{v}\rVert_{0}\leq s$
where $1\le s \le p$, the optimal solution is $
\boldsymbol{v}^{\star}= {\bsbTh^{\#}(\boldsymbol{y};s)}/{\lVert \bsbTh^{\#}(\boldsymbol{y};s)\rVert_{2}}. $
\end{lemma}
Here, $\Theta^\#$ is the quantile thresholding \citep{She2017}. 
The lemma can be used to calculate the sparse regression depth in  \cite{chen2018robust}. 
\begin{proof}
Let $\mathcal{J}=\{j:v_{j}\neq0\}$, $\mathcal{J}^{c}=\{j:v_{j}=0\}$
and $\mathcal{V}(\mathcal{J})=\{\boldsymbol{v}\in\mathbb{R}^{p}:v_{j}=0\;\textrm{for}\;j\in\mathcal{J}^{c}\}$.
Given $\mathcal{J}$, the optimal solution
of
\[
\min_{\boldsymbol{v}\in\mathcal{V}(\mathcal{J})}\lVert\boldsymbol{y}-\boldsymbol{v}\rVert_{2}^{2}\;\textrm{s.t.}\;\lVert\boldsymbol{v}\rVert_{2}=1
\]
is $\boldsymbol{v}_{\mathcal{J}}^{\star}=\boldsymbol{y}_{\mathcal{J}}/\lVert\boldsymbol{y}_{\mathcal{J}}\rVert_{2}$
and $\boldsymbol{v}_{\mathcal{J}^{c}}=\boldsymbol{0}$.   The problem thus reduces to
\[
\min_{\mathcal J:\lvert\mathcal{J}\rvert\leq s}\lVert\boldsymbol{y}_{\mathcal{J}^{c}}\rVert_{2}^{2}+\lVert\boldsymbol{y}_{\mathcal{J}}-\boldsymbol{v}_{\mathcal{J}}^{\star}\rVert_{2}^{2},
\]
or
\[
\min_{\lvert\mathcal{J}\rvert\leq s}\lVert\boldsymbol{y}_{\mathcal{J}^{c}}\rVert_{2}^{2}+(\lVert\boldsymbol{y}_{\mathcal{J}}\rVert_{2}-1)^{2}.
\]
Noticing that
\begin{align*}
 & \lVert\boldsymbol{y}_{\mathcal{J}^{c}}\rVert_{2}^{2}+(\lVert\boldsymbol{y}_{\mathcal{J}}\rVert_{2}-1)^{2}\\
= & \lVert\boldsymbol{y}_{\mathcal{J}^{c}}\rVert_{2}^{2}+\lVert\boldsymbol{y}_{\mathcal{J}}\rVert_{2}^{2}-2\lVert\boldsymbol{y}_{\mathcal{J}}\rVert_{2}+1\\
= & \lVert\boldsymbol{y}\rVert_{2}^{2}+1-2\lVert\boldsymbol{y}_{\mathcal{J}}\rVert_{2},
\end{align*}
the conclusion follows.
\end{proof}

\section{Proof of Theorem \ref{thm:gradient of f Lip continous}}

By the construction of $g$ and the definition of $\bsbV^{(t+1)}$, we have
\[
g_{\rho_{t}}(\boldsymbol{V}^{(t+1)},\boldsymbol{V}^{(t)})\leq g_{\rho_{t}}(\boldsymbol{V}^{(t)},\boldsymbol{V}^{(t)})=f(\boldsymbol{V}^{(t)}).
\]
It  remains to show $f(\boldsymbol{V}^{(t+1)})\leq g_{\rho_{t}}(\boldsymbol{V}^{(t+1)},\boldsymbol{V}^{(t)})$. We prove a stronger result: for any  $\bsbV, \bsbV^{-}\in \mathbb R^{p\times m}$,
\[
f ( \boldsymbol { V } ) - g _ { \rho   }  ( \boldsymbol { V } , \boldsymbol { V } ^ { - }  ) =f(\boldsymbol{V})-f(\boldsymbol{V}^{-})-\langle\nabla f(\boldsymbol{V}^{-}),\boldsymbol{V}-\boldsymbol{V}^{-}\rangle-\frac{\rho}{2}\lVert\boldsymbol{V}-\boldsymbol{V}^{-}\rVert_{F}^{2}\leq0 \]
provided that $\rho \ge L \| \bsbX\|_2^2 \| \bsbR\|_2^2$. In fact,
\begin{align}
 & f(\boldsymbol{V})-f(\boldsymbol{V}^{-})-\langle\nabla f(\boldsymbol{V}^{-}),\boldsymbol{V}-\boldsymbol{V}^{-}\rangle\notag\\
= & \int_{0}^{1}\langle\nabla f(\boldsymbol{V}^{-}+t(\boldsymbol{V}-\boldsymbol{V}^{-})),\boldsymbol{V}-\boldsymbol{V}^{-}\rangle \rd t-\int_{0}^{1}\langle\nabla f(\boldsymbol{V}^{-}),\boldsymbol{V}-\boldsymbol{V}^{-}\rangle \rd t\notag\\
= & \int_{0}^{1}\langle\nabla f(\boldsymbol{V}^{-}+t(\boldsymbol{V}-\boldsymbol{V}^{-}))-\nabla f(\boldsymbol{V}^{-}),\boldsymbol{V}-\boldsymbol{V}^{-}\rangle \rd t\notag\\
\leq & \int_{0}^{1}\lVert\nabla f(\boldsymbol{V}^{-}+t(\boldsymbol{V}-\boldsymbol{V}^{-}))-\nabla f(\boldsymbol{V}^{-})\rVert_{F}\lVert\boldsymbol{V}-\boldsymbol{V}^{-}\rVert_{F} \rd t. \label{intbound}
\end{align}

It is easy to verify that   $\nabla f(\boldsymbol{V})= \sum_i \bsbx_i \varphi'(\bsbx_i^T \bsbV \bsbr_i) \bsbr_i^T=\boldsymbol{X}^{T}\textrm{diag}(\varphi^{\prime}(\boldsymbol{X}\boldsymbol{V}\boldsymbol{R}^{T}))\boldsymbol{R}$.  Given any $\bsbV, \bsbV^-$,
\begin{align*}
& \lVert\nabla f(\boldsymbol{V})-\nabla f(\boldsymbol{V}^{-})\rVert_{F} \\
  =&\rVert\boldsymbol{X}^{T}\{\textrm{diag}(\varphi^{\prime}(\boldsymbol{X}\boldsymbol{V}\boldsymbol{R}^{T}))-\textrm{diag}(\varphi^{\prime}(\boldsymbol{X}\boldsymbol{V}^{-}\boldsymbol{R}^{T}))\}\boldsymbol{R}\rVert_{F}\\
 =&\rVert (\boldsymbol{R}^T \otimes \boldsymbol{X}^{T}) \vect( \textrm{diag}(\varphi^{\prime}(\boldsymbol{X}\boldsymbol{V}\boldsymbol{R}^{T})-\varphi^{\prime}(\boldsymbol{X}\boldsymbol{V}^{-}\boldsymbol{R}^{T})))\rVert_{2}\\
   \le & \| \boldsymbol{R}^T \otimes \boldsymbol{X}^{T}\|_2 \times \|  \textrm{diag}(\varphi^{\prime}(\boldsymbol{X}\boldsymbol{V}\boldsymbol{R}^{T})-\varphi^{\prime}(\boldsymbol{X}\boldsymbol{V}^{-}\boldsymbol{R}^{T}))\|_F \\
  \le & L \| \bsbX\|_2 \| \bsbR\|_2 \| \mbox{diag}(\boldsymbol{X}\boldsymbol{V}\boldsymbol{R}^{T} - \boldsymbol{X}\boldsymbol{V}^{-}\boldsymbol{R}^{T})\|_F\\
 \le &L \| \bsbX\|_2 \| \bsbR\|_2 \|\boldsymbol{X}\boldsymbol{V}\boldsymbol{R}^{T} -  \boldsymbol{X}\boldsymbol{V}^{-}\boldsymbol{R}^{T} \|_F \\
  \le &  L \| \bsbX\|_2 \| \bsbR\|_2 \| \bsbR \otimes \bsbX \|_2 \| \bsbV - \bsbV^{-}\|_F \\
 = &L \|\bsbX\|_2^2 \| \bsbR\|_2^2 \|\bsbV - \bsbV^{-}\|_F,
\end{align*}
 where we used $\vect (\bsbA \bsbX \bsbB) = (\bsbB^T \otimes \bsbA )\vect (\bsbX)$  and $\| \bsbA \otimes \bsbB\|_2 = \| \bsbA\|_2 \| \bsbB\|_2$ twice, together with the assumption on $\varphi$.
(A finer bound can be given:
$\lVert\nabla f(\boldsymbol{V})-\nabla f(\boldsymbol{V}^{-})\rVert_{F} \le L \| \bsbX\|_2 \| \bsbR\|_2  \| \mbox{diag}(\boldsymbol{X}\boldsymbol{V}\boldsymbol{R}^{T} - \boldsymbol{X}\boldsymbol{V}^{-}\boldsymbol{R}^{T})\|_F\le L \| \bsbX\|_2 \allowbreak  \| \bsbR\|_2(\sum \|\bsbx_i\|_2^2 \|\bsbr_i\|_2^2)^{1/2}\|\bsbV - \bsbV^{-}\|_F$.)

Plugging this result into \eqref{intbound}, we get
\begin{align*}
& f(\boldsymbol{V})-f(\boldsymbol{V}^{-})-\langle\nabla f(\boldsymbol{V}^{-}),\boldsymbol{V}-\boldsymbol{V}^{-}\rangle
\\
\leq & \,  \int_{0}^{1}  L \|\bsbX\|_2^2 \| \bsbR\|_2^2 \lVert t(\boldsymbol{V}-\boldsymbol{V}^{-})\rVert_{F}\lVert\boldsymbol{V}-\boldsymbol{V}^{-}\rVert_{F}  \rd t.
 \\ = & \, L \|\bsbX\|_2^2 \| \bsbR\|_2^2\int_{0}^{1} \lVert\boldsymbol{V}-\boldsymbol{V}^{-}\rVert_{F}^{2}t \rd t \\
= \, & \frac{L \|\bsbX\|_2^2 \| \bsbR\|_2^2}{2}\lVert\boldsymbol{V}-\boldsymbol{V}^{-}\rVert_{F}^{2}.
\end{align*}
The conclusion follows.
\section{Proof of Theorem \ref{th:comp_acc2}}

It is not difficult to see that $\bsbW^{(t+1)}$ solves $\min_{\bsbV} \| \bsbXi^{(t)} -\bsbV\|_F  $ s.t. $\bsbV \in \mathcal G, \|\bsbV\|_F=1$, and is thus  a globally optimal solution to $$\min_{\bsbV}f(\bsbV) - \bm\Delta_{f}(\bsbV,\bsbU^{(t)}) + \theta_t\rho_t\mathbf{D}_2 (\bsbV,\bsbW^{(t)}) \  \mbox{ subject to } \  \bsbV \in \mathcal G, \|\bsbV\|_F=1.$$

\begin{lemma}\label{pertubres0}
Let $l(\bsbv) = (1/2) \| \bsbv - \bsby \|_2^2$ and $\bsbv_o$ be $ \bsby / \|\bsby\|_2$ if $\bsby   \ne  \bsb0$ and an arbitrary unit vector otherwise. Then for any $\bsbv: \bsbv^T \bsbv = 1$, $l(\bsbv) - l(\bsbv_o)  = \| \bsby \|_2 \|\bsbv_o - \bsbv \|_2^2/2$.
\end{lemma}
The proof is simple and omitted.

For convenience, we denote $l_{f}(\bsbV,\bsbU) = f(\bsbV) - \bm\Delta_{f}(\bsbV,\bsbU)$.
According to Lemma \ref{pertubres0},  for any $ \bsbV \in \mathcal G: \|\bsbV\|_F=1$ we have
\begin{equation} \label{tri_acc2}
\begin{split}
&l_{f}(\bsbW^{(t+1)},\bsbU^{(t)}) - l_{f}(\bsbV,\bsbU^{(t)}) + \theta_t\rho_t\mathbf{D}_2 (\bsbW^{(t+1)},\bsbW^{(t)})\\
\le\,& \theta_t\rho_t\mathbf{D}_2 (\bsbV,\bsbW^{(t)}) -   \theta_t\rho_t \| \bsbXi^{(t)}\|_F\mathbf{D}_2   (\bsbV, \bsbW^{(t+1)}).
\end{split}
\end{equation}
By the linearity of  $l_f(\cdot, \bsbU^{(t)})$,
\begin{equation} \label{tri_acc2-1}
\begin{split}
0 = \theta_t l_{f}(\bsbW^{(t+1)},\bsbU^{(t)}) + (1-\theta_t) l_{f}(\bsbV^{(t)},\bsbU^{(t)}) - l_{f}(\bsbV^{(t+1)},\bsbU^{(t)}).
\end{split}
\end{equation}
Multiplying \eqref{tri_acc2} by $\theta_t$, and adding the resultant inequality to \eqref{tri_acc2-1},  we obtain
\begin{equation}\nonumber
\begin{split}
&l_{f}(\bsbV^{(t+1)},\bsbU^{(t)}) - (1-\theta_t)l_{f}(\bsbV^{(t)},\bsbU^{(t)}) - \theta_t l_{f}(\bsbV,\bsbU^{(t)})\\
&       + \theta_t^2\rho_t\mathbf{D}_2 (\bsbW^{(t+1)},\bsbW^{(t)})+ \theta_t^2\rho_t (\| \bsbXi^{(t)}\|_F-1)\mathbf{D}_2   (\bsbV, \bsbW^{(t+1)})\\
\le\,&\theta_t^2\rho_t\mathbf{D}_2 (\bsbV,\bsbW^{(t)}) -  \theta_t^2\rho_t\mathbf{D}_2  (\bsbV,\bsbW^{(t+1)}),
\end{split}
\end{equation}
and so
\begin{equation}\label{intermdres1}
\begin{split}
&f(\bsbV^{(t+1)})-f(\bsbV) - (1-\theta_t)(f(\bsbV^{(t)})-f(\bsbV)) + R_t \\
&+ \theta_t\{\bm\Delta_{f}(\bsbV,\bsbU^{(t)}) +  \theta_t\rho_t (\| \bsbXi^{(t)}\|_F-1)\mathbf{D}_2   (\bsbV, \bsbW^{(t+1)})
\} \\
\le\,&\theta_t^2\rho_t(\mathbf{D}_2 (\bsbV,\bsbW^{(t)}) - \mathbf{D}_2 (\bsbV,\bsbW^{(t+1)})),
\end{split}
\end{equation}
where $R_t = \theta_t^2\rho_t\mathbf{D}_2 (\bsbW^{(t+1)},\bsbW^{(t)}) - \bm\Delta_{f}(\bsbV^{(t+1)},\bsbU^{(t)}) + (1-\theta_t)\bm\Delta_{f}(\bsbV^{(t)},\bsbU^{(t)}) $.
We rewrite \eqref{intermdres1} into the following recursive form
\begin{equation} \label{eq:th_acc2-4}
\begin{split}
&\frac{1}{\theta_t^2\rho_t}\big[f(\bsbV^{(t+1)})-f(\bsbV)\big] - \frac{1-\theta_t}{\theta_{t}^2\rho_{t}}\big[f(\bsbV^{(t)})-f(\bsbV)\big] + \frac{\mathcal E_t(\bsbV)}{\theta_t\rho_t} + \frac{R_t}{\theta_t^2\rho_t}\\
\le\,&\mathbf{D}_2 (\bsbV,\bsbW^{(t)}) - \mathbf{D}_2 (\bsbV,\bsbW^{(t+1)})
\end{split}
\end{equation}
with $\mathcal E_t(\bsbV) = \bm\Delta_{f}(\bsbV,\bsbU^{(t)}) +   \theta_t\rho_t( \| \bsbXi^{(t)}\|_F-1)\mathbf{D}_2   (\bsbV, \bsbW^{(t+1)})
$.
It follows from \eqref{acc2_search2} that
\begin{equation} \label{eq:th_acc2-5}
\begin{split}
&\frac{1}{\theta_t^2\rho_t}\big[f(\bsbV^{(t+1)})-f(\bsbV)\big] - \frac{1}{\theta_{t-1}^2 \rho_{t-1}}\big[f(\bsbV^{(t)})-f(\bsbV)\big] + \frac{\mathcal E_t(\bsbV)}{\theta_t\rho_t} + \frac{R_t}{\theta_t^2\rho_t}\\
\le\,&\mathbf{D}_2 (\bsbV,\bsbW^{(t)}) - \mathbf{D}_2 (\bsbV,\bsbW^{(t+1)}).
\end{split}
\end{equation}
Applying \eqref{eq:th_acc2-5} with $t= T, \ldots, 1$, and \eqref{eq:th_acc2-4} with $t=0$, and adding all inequalities together, we have
\begin{equation} \nonumber
\begin{split}
&\frac{1}{\theta_T^2\rho_T}\big[f(\bsbV^{(T+1)})-f(\bsbV)\big] - \frac{1-\theta_0}{\theta_0^2\rho_0}\big[f(\bsbV^{(0)})-f(\bsbV)\big] + \sum_{t=0}^T \Big(\frac{\mathcal E_t(\bsbV)}{\theta_t\rho_t} + \frac{R_t}{\theta_t^2\rho_t}\Big)\\
\le\,&\mathbf{D}_2 (\bsbV,\bsbW^{(0)}) - \mathbf{D}_2 (\bsbV,\bsbW^{(T+1)}).
\end{split}
\end{equation}
Noticing that $\theta_0=1$, we obtain the conclusion  from the following result
\begin{equation} \nonumber
\frac{1}{\theta_T^2\rho_T}\big[f(\bsbV^{(T+1)})-f(\bsbV)\big] + \sum_{t=0}^T\Big(\frac{ \mathcal E_t(\bsbV)}{\theta_t\rho_t} + \frac{R_t}{\theta_t^2\rho_t}\Big) \le \mathbf{D}_2 (\bsbV,\bsbW^{(0)}) - \mathbf{D}_2 (\bsbV,\bsbW^{(T+1)}).
\end{equation}
which holds for any $\bsbV \in \mathcal G: \|\bsbV\|_F=1$.

{
\bibliographystyle{apalike}
\bibliography{DataDepth}
}


\end{document}